\documentclass[3p, 10pt]{elsarticle}
\usepackage{amssymb, amsmath}
\usepackage{amsthm}
\usepackage{subfig}
\usepackage{bm}
\usepackage{units}
\setcitestyle{authoryear}
\setcitestyle{citesep={;}}
\usepackage{xcolor}
\usepackage{capt-of}
\usepackage{savesym}
\savesymbol{AND}

\usepackage{algorithmic}
\usepackage{algorithm}
\usepackage{setspace}
\usepackage{hyperref}

\theoremstyle{definition}

\theoremstyle{remark}

\usepackage[section]{placeins}
\usepackage{lineno}
\renewcommand{\linenumberfont}{\normalfont\footnotesize\color{gray}}

\makeatletter
\def\makeLineNumberLeft{%
  \linenumberfont\llap{\hb@xt@\linenumberwidth{\LineNumber\hss}\hskip\linenumbersep}% left line
  \hskip\columnwidth% skip over column of text
  \rlap{\hskip\linenumbersep\hb@xt@\linenumberwidth{\hss\LineNumber}}\hss}% right line number
\leftlinenumbers% Re-issue [left] option
\makeatother

\renewcommand{\L}{\ensuremath{\mathcal{L}}}
\newcommand{\T}{\ensuremath{\mathsf{T}}}
\renewcommand{\b}{\boldsymbol}
\newcommand{\x}{\b{x}}

\hypersetup{
  colorlinks, linkcolor=blue
}
\begin{document}
\onecolumn \vspace*{-2.5cm} 
%TITLE
\begin{center}
  \textbf{\Large \noindent RBF-FD analysis of 2D time-domain acoustic wave propagation in
  heterogeneous media \vspace{0.2cm}} \vspace{2cm}\\
  Jure Močnik - Berljavac$^{a,b}$,  Pankaj K Mishra$^{c} $,
  Jure Slak$^{a,b}$ and Gregor Kosec$^{b} $\vspace{1cm}\\
  $^a$ \textit{\small Faculty of Mathematics and Physics, University of Ljubljana, Jadranska 19,
  1000 Ljubljana, Slovenia  }\\
  $^b$ \textit{\small ``Jožef Stefan'' Institute, Department E6, Parallel and Distributed Systems
  Laboratory, Jamova cesta 39, 1000 Ljubljana, Slovenia}\\
  $^c$ \textit{\small Institute for Geophysics, The University of Texas at Austin, 10100 Burnet
  Road, Austin, Texas, USA}\\
\end{center}
\vspace{1cm}

%=======================================================================
\section*{Abstract}
\noindent\makebox[\linewidth]{\rule{16.5cm}{0.4pt}}
Radial Basis Function-generated  Finite Differences (RBF-FD) is a popular variant of
local strong-form meshless methods that do not require a predefined connection between the nodes,
making it easier to adapt node-distribution to the problem under consideration. This paper
investigates an RBF-FD solution of time-domain acoustic wave propagation in the context of seismic
modeling in the Earth's subsurface. Through a number of numerical tests, ranging from homogeneous
to highly-heterogeneous velocity models including non-smooth irregular topography, we demonstrate that the present approach can be further generalized to solve large-scale seismic modeling and full waveform inversion problems in arbitrarily complex models enabling more robust interpretations of geophysical observations.

\noindent\makebox[\linewidth]{\rule{16.5cm}{0.4pt}}
\section{Introduction}
Numerical modelling is a widely used approach for
computational simulation of geological processes. Numerical approximation of acoustic wave
equation in complex velocity media is vital to a wide range of investigations
in geophysics
seismic modelling, reverse-time migration, seismic inversion, etc. To simulate
the acoustic waves in a complex representation of the Earth's subsurface, time-domain wave
equation is often solved approximately, using mesh or
grids to discretize the domain of interest. Over the
years, a wide range of numerical methods have been proposed and applied for
acoustic wave
simulations in geoscience, including Finite Difference Method
(\cite{alford1974accuracy,kelly1976synthetic,tarantola1984inversion,dablain1986application,
  williamson1995critical,jo1996optimal,carcione2002seismic,geiger,du20042,liu2011finite,Virieux2012,
  wang2016effective,wang2018time,wang2019optimized,cai2018acoustic}),
 Finite Element Method
(\cite{marfurt1984accuracy,emmerich1987incorporation,de2007grid,ham2012finite}), Spectral Element
Method (\cite{seriani1994spectral,seriani2007optimal,shukla2019modeling,malovichko2018acoustic}). Finite difference method (FDM) has been
frequently preferred over other methods, due to its excellent compromise between accuracy,
stability, and computational efficiency. Nevertheless, FDM has its shortcomings. Given the
complexity of the Earth model, it is often desirable to use spatially variable discretization,
which could  potentially also be adaptive to the velocity variations
\cite{jastram1992acoustic,hayashi2001discontinuous,kang2004efficient,
  kristek2010stable,chu2012nonuniform}.
 FDM does not offer such flexibility, at least not without special treatment.

However, the Radial Basis Function Generated Finite Differences (RBF-FD)
method~\cite{fornberg1988generation}, a generalization of FDM, do not require a predefined
grid, and therefore offers great flexibility regarding the geometry and of the
domain as well as the distribution of nodes. The conceptual difference between FDM and RBF-FD is in the way the nodes are treated. FDM uses a priori knowledge about the nodes and their connectivity with
neighbours, as the nodes are organized in a grid that is known in advance. In
RBF-FD no a priori knowledge about the nodal topology is required and the support domains are defined
in the solution procedure, but at a larger cost to memory, since generally each node has a different 
local neighbourhood. A direct consequence of higher flexibility regarding the nodal positioning is
that RBF-FD is, in contrast to FDM, able to locally modify node configurations by simply placing more 
points in areas where needed and removing them from areas that are already 
overpopulated~\cite{slak2019adaptive}. The RBF-FD method is a
popular variant out of many strong-form local meshless methods. It uses finite difference-like 
collocation weights on an unstructured set of nodes~\cite{tolstykh2003using}. The method has been 
successfully used in several problems and is still actively researched~\cite{Fornberg_Flyer_2015, bayona2017role,slak2018refined,mishra2019stabilized,slak2019generation}.

Previous works for modeling acoustic wave equations using weak-form meshfree methods
include~\cite{jia2005meshless,hahn2009use,zhang2016efficient} and using strong-form meshfree
methods include ~\citep{takekawa2015mesh,takekawa2016mesh,liu2017perfectly,mishra2017frequency,takekawa2018mesh}.
The strong-form meshfree investigations, mentioned above, implement meshfree computations only in
the space-domain (frequency-domain approximation of the acoustic wave equation). Recently,
\cite{li2017time} presented a first investigation of application of a mesh-free FD method, based
on least squares optimization, for time-domain simulation of acoustic wave equation.
Motivated by the success and robustness of RBF-FD
\cite{Fornberg_Flyer_2015,fornberg1988generation, slak2018refined, slak2019generation}, it is
intriguing to test them on an extended spectra of problems. In this
paper, we
present an investigation of RBF-FD method for modelling 2D time-domain acoustic
wave propagation
in heterogeneous Earth's subsurface.
In order to suppress the artificial reflections arising from the
truncation of the computational domain while mimicking the infinitely large-domain, we couple
absorbing boundary conditions with the RBF-FD formulation.

The rest of the paper is structured as follows. In section~\ref{sec:rbffd}, we discuss the
general RBF-FD formulation for solving PDEs and different aspects of its successful implementation.
In section~\ref{sec:model}, we explain the governing equations of the time-domain acoustic wave
propagation and the absorbing boundary conditions. In section~\ref{sec:num}, a series of numerical
tests for modelling the wave propagation in (1) homogeneous (2) layered, and (3)
highly-heterogeneous Marmousi velocity model of the subsurface have been performed. Standard FD
results are provided in first two cases for a heuristic comparison. All examples were computed
using the in-house~\cite{Medusa} library. This is followed by the conclusions and some potential
future works.

\section{RBF-FD formulation}
\label{sec:rbffd}
RBF-FD, as the name suggests, is a generalization of the Finite Difference Method (FDM). Both
methods use computational nodes, or points, at which the solution is approximated. Both are also
local, meaning only nodes 'close' to the selected node can affect the selected node's next value.
This neighbourhood of close nodes is commonly referred to as a stencil or the
support domain.

Classical FDM approximates differential operators with a weighted combination of
neighbouring nodal values, for example
\begin{equation}
  u''(x_i) \approx
  \frac{1}{h^2}u(x_{i-1})
   -\frac{2}{h^2} u(x_{i})
  + \frac{1}{h^2}u(x_{i+1})  =
   \begin{bmatrix} \frac{1}{h^2} & -\frac{2}{h^2} & \frac{1}{h^2} \end{bmatrix}
  \begin{bmatrix}u(x_{i-1}) \\ u(x_{i})\\ u(x_{i+1}) \end{bmatrix}
\end{equation}
for second derivatives in 1D. We can compute and use
$[1/h^2, -2/h^2, 1/h^2]$ as an approximation for the second derivative,
evaluated at a centre point irrespective of the actual function values.
RBF-FD uses the same methodology in a more general setting, were such weights cannot be precomputed,
and their computation is considered a part of the solution procedure.

For a general partial differential operator $\L$ at a point $\x_i$, we seek the
approximation
in the form
\begin{equation} \label{eq:approx}
  (\L u)(\x) \approx \sum_{\x_j \in S(\x)} w_{j}(\x) u(\x_j) = \b w(\x)^\T \b u,
\end{equation}
where $S(\x)$ represents the neighbouring nodes (also called
\emph{stencil} or \emph{support}) of $\x$. Denote the number of neighbours with $n = |S(\x)|$.

To compute the weights, we write $n$ linear equation, obtained from enforcing exactness
of~\eqref{eq:approx} for a class of functions. In RBF-FD method, these are
radial basis functions, centred in stencil nodes. Many different choices for RBFs exists;
we sill use the Gaussians, defined as
\begin{align}
    \Phi_{\x_j}(\x) = \phi(\|\x-\x_j|), \quad \phi(r)= \exp(-r^2/\sigma_B^2),
\end{align}
where $\sigma_B$ is a positive real shape parameter.
Substituting $\Phi_{\x_k}$ for all $\x_k \in S$ in place of $u$ in~\ref{eq:approx}
gives rise to a system of $n$ linear equations
\begin{equation}
\label{eq:system}
\begin{bmatrix}
\Phi_{\x_1}(\x_1) & \cdots & \Phi_{\x_1}(\x_n) \\
\vdots & \ddots & \vdots \\
\Phi_{\x_n}(\x_1) & \cdots & \Phi_{\x_n}(\x_n) \\
\end{bmatrix}
\begin{bmatrix}
w_1 \\ \vdots \\ w_n
\end{bmatrix} =
\begin{bmatrix}
(\L \Phi_{\x_1})(\x) \\
\vdots \\
(\L \Phi_{\x_n})(\x) \\
\end{bmatrix},
\end{equation}
which can be compactly written as $A_\phi \b w(\x) = \b \ell$ and
solved to obtain $\b w(\x)$.
The matrix $A_\phi$ is symmetric and when Gaussian basis
functions are used, it is also positive definite~\cite{Fornberg_Flyer_2015}. This guaranties
non-singularity as long as all support domain nodes are distinct.

To obtain the solution of a PDE, we first discretise the domain $\Omega$ and its boundary
$\partial \Omega$  with $N$ nodes. For each computational node $\x_i$
we compute the stencil $S(\x_i)$, consisting of its $n$ closest neighbours.
Then, we compute and store the weights $\b w(\x_i)$ for all nodes $x_i$ and all operators
$\L$ in the equation, including possible differential operators used to define boundary
conditions, such as normal derivatives. Since the nodes do not change during the simulation,
computed values can be stored and used to effectively obtain approximations
of field derivatives in $O(n)$ time by using a simple dot product, as
posed in~\eqref{eq:approx}.

In all numerical examples we will use collocation with $m=7$ Gaussian functions on supports of $n=7$ closest nodes. A Poisson Disk Sampling-based node generation
algorithm~\cite{slak2019generation} will be used to position the nodes. The
algorithm strives to position nodes as regular as possible in an arbitrary
domain with a supplied spatially dependent target distance between nodes,
effectively enabling the ability to refine the numerical solution~\cite{slak2018refined}.
The weights $\b w_i$ for the Laplacian operator $\L = \frac{\partial^2}{\partial x^2} + \frac{\partial^2}{\partial z^2}$ are computed in advance for all interior nodes $\x_i$, using~\eqref{eq:system} and stored, to approximate the spatial part of the equation as
\begin{equation}
    \frac{\partial^2u}{\partial x^2}(x_i, z_i, t) + \frac{\partial^2u}{\partial z^2}(x_i, z_i, t) \approx \sum_{j \in S(\x_i)} (\b w_i)_j u_j(t) =: \b w_i^{\mathsf T} \cdot \b u_{S(\x_i)},
\end{equation}
where values $u_j$ represent the function values in the computational nodes $\x_j$ and 
$u_{S(\x_i)}$ represent the subset of $u_j$ that correspond to the neighbours of $\x_i$.

Explicit time stepping is used for time discretization
\begin{equation}
    \frac{d^2u}{dt^2} \approx \frac{u^{(n-2)} - 2 u^{(n-1)} + u^{(n)} } {\Delta t^2},
\end{equation}
where $u^{(n-2)}$ and $u^{(n-1)}$ stand for previous two time steps and $u^{(n)}$ for the current time step. Initially all fields are set to zero.

\section{Model of acoustic wave propagation in the Earth}
\label{sec:model}
% Seismic waves induce elastic deformation while propagating through the
% Earth's subsurface, which can be recorded and used to interpret the subsurface structure. The
% equation of wave motion, representing the general description of the medium, are derived by using
% stress-strain relationships (the Hooke’s law) and momentum
% equations~\cite{shearer2019introduction}. In order to simulate a realistic geophysical scenario,
% one has to solve 3D seismic wave equation, which is often computationally expensive ---
% especially in the inverse problems where the wave equations needs to be solved at every
% iteration. This becomes more problematic when using global optimization algorithms where the
% wave-equation is solved for thousands of iterations. Therefore, it is often more practical to
%solve a constant-density acoustic approximation of the seismic wave propagation~\cite{biswas20172d}.
The standard 2D constant-density approximation of the time-domain acoustic wave equation is
given as
\begin{equation}
\frac{1}{v_p(x,z)^2}\frac{\partial^2 u(x,z,t)}{\partial t^2} = \frac{\partial^2
	u(x,z,t)}{\partial x^2}+\frac{\partial^2 u(x,z,t)}{\partial z^2} +\delta(x - x_s, z-z_s)s(t),
\end{equation}
where $u$ is the pressure amplitude or pressure wavefield and $v_p(x,z)$ is primary wave (P-wave)
velocity, which represents the material properties of the subsurface.

In general, the domain of interest is the entire subsurface of the Earth, which can
from local
point of view be seen as
\begin{equation}
\Omega = \{(x,z,t) | -\infty <x<+\infty, -\infty <z<+\infty, t\geq 0 \}.
\end{equation}
However, practical computational limitations enforce a constraint on the size of the domain. Therefore
the actual computational domain is represented as
\begin{equation}
\Omega = \{(x,z,t) | x_{min} <x< x_{max}, 0 <z< z_{max}, t\geq 0 \},
\end{equation}
with Dirichlet boundary conditions on all sides. Since infinite space is represented
through finite computational domain, the reflections from the boundaries are undesired and called
as spurious reflections. There are a number of approaches to suppress such spurious reflections
from the numerical solution, out of which, we choose one of the most simple formulation termed as
``Absorbing boundary conditions (ABC)'' proposed by \cite{cerjan1985}. The idea behind ABC is to
introduce a spatially variable damping factor, which starts at a given distance from the boundary
and increases its weight as it approaches the boundary being maximum at the boundary. The damping
factor is given by
\begin{equation}
\label{eq:abc}
G(i) = \exp \left(-\left[0.015(i_{max}-i)\right]^2\right),
\end{equation}
where $i_{max}$ is the thickness of the absorbing layer in terms of nodes, that is, the number of
nodes along the thickness of the absorbing layer. This damping factor is multiplied to the
wavefield, which, practically, reduces its amplitude to zero at the boundary suppressing any
undesired reflections from that boundary.

When using RBF-FD the nodes can be scattered and slight modification need to be made to the ABCs.
Due to the irregular node layout, a
continuous form of~\eqref{eq:abc} is needed, given as
\begin{equation}
\label{eq:abs_con_disc2}
G(d) = \exp \left(-\left[0.015(i_{max} - d/h)\right]^2\right),
\end{equation}
where $d$ represents the distance from the current node to the boundary and $h$ is the
current average nodal spacing.

The remaining boundary condition is the top boundary at $z=0$, which
represents the Earth's surface. The reflections from this boundary are of physical origin,
there is no need for the absorbing layer and ordinary Dirichlet conditions suffice.

\begin{figure}[t]
  \centering
  \subfloat{
  \includegraphics[width=0.5\textwidth]{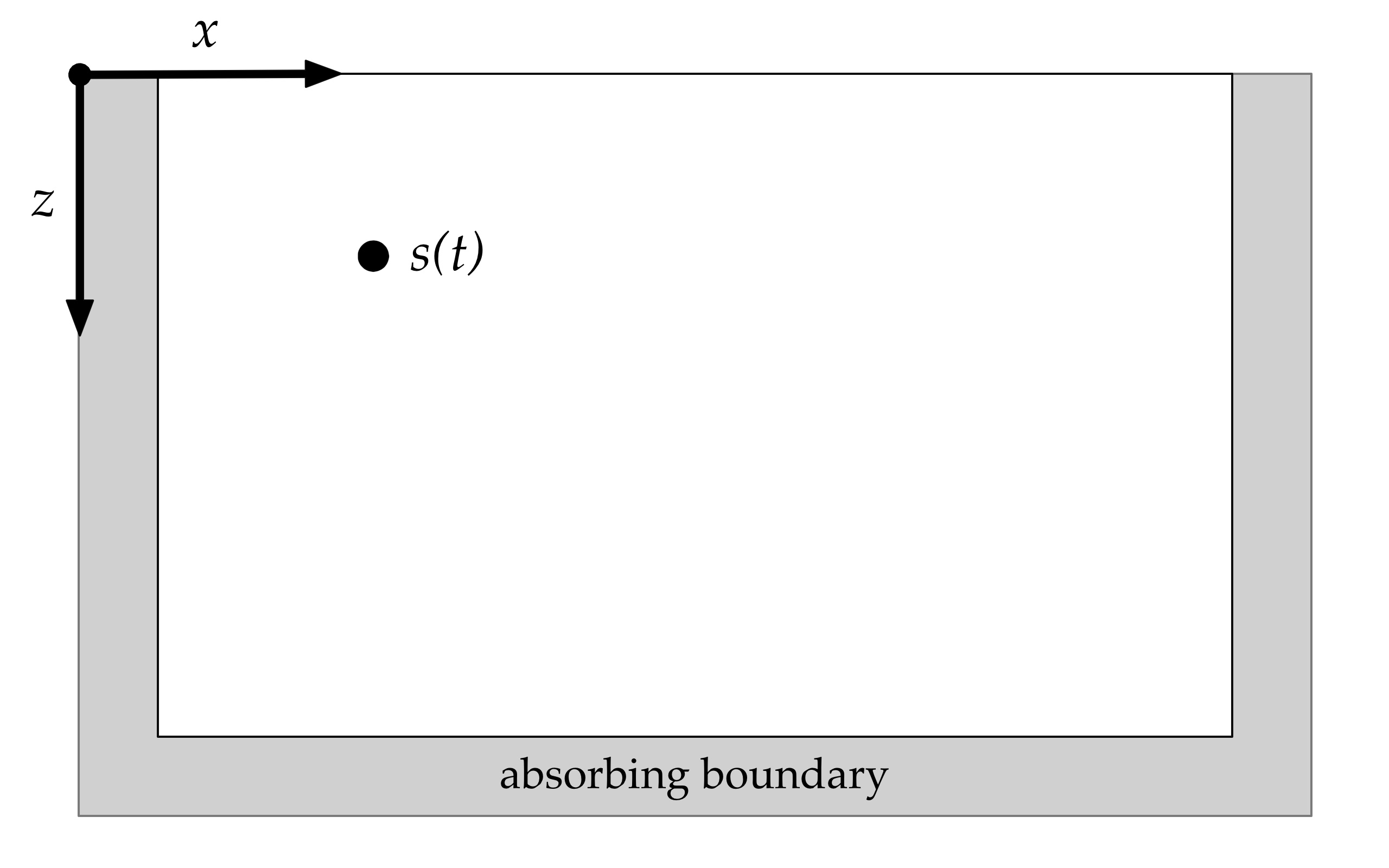}
  }
  \subfloat{
    \includegraphics[width=.41\textwidth]{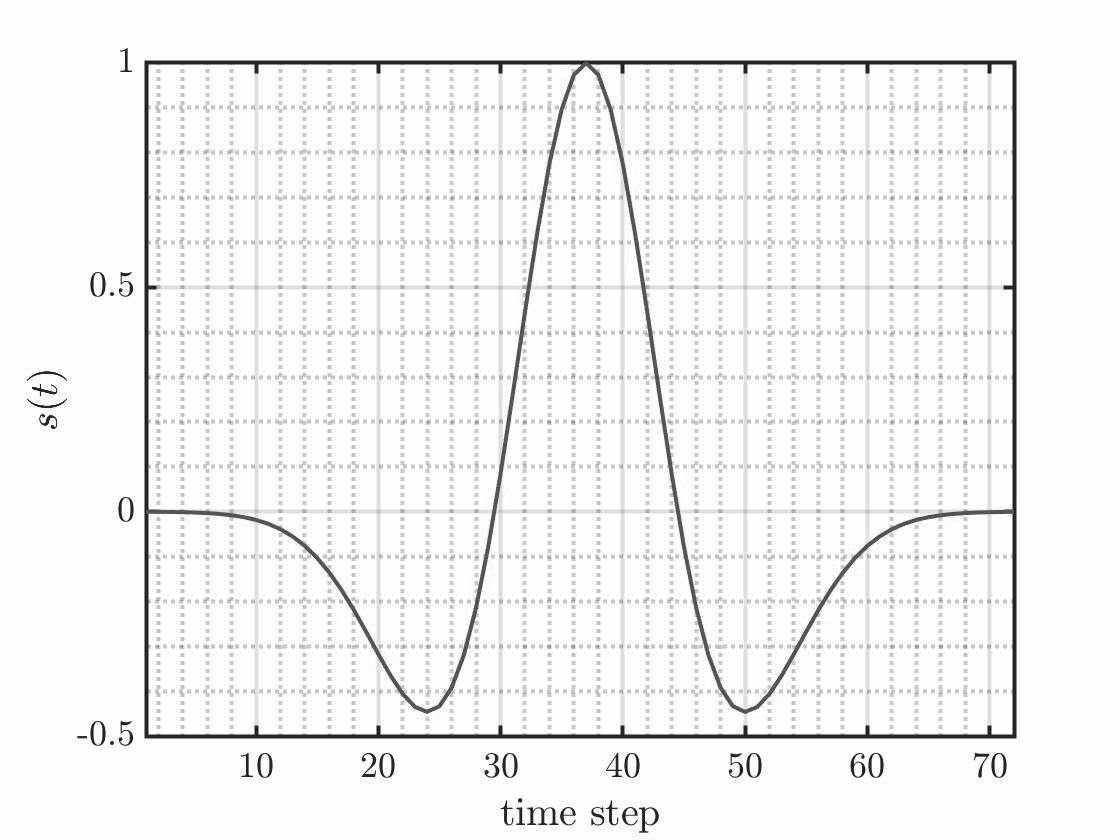}
  }
  \caption{Domain of interest with absorbing boundary layers (left) and
  Ricker's wavelet; $\sigma_R=7.5$ (right).}

  \label{fig:domaininfo}
\end{figure}

The wave source is given as Ricker's Wavelet, shown in Figure~\ref{fig:domaininfo}.
It is formally given by
\begin{equation}
\label{eq:ricker}
s(t)= \frac{2 s_0}{\sqrt{3\sigma_R}\pi^{1/4}}\Big(1-\Big(\frac{t}{\sigma_R}\Big)^{\!\!2}\Big)
e^{-\frac{t^2}{2\sigma_R^2}},
\end{equation}
where $\sigma_R$ is the shape parameter and $s_0$ is the amplitude.
The wave source also includes a $\delta$-function, which is implemented as
\begin{equation}
\delta(x, z)\simeq\frac{1}{\pi}\frac{\epsilon}{x^2+z^2+\epsilon^2}.
\end{equation}
where $\epsilon$ is a small constant,
larger than the nodal spacing $h$, so that the source can be
adequately represented regardless of the current discretization.
We will use the value $\epsilon=\unit[4.0]{m}$ in our paper.

%=====================================================
\section{Numerical examples}
\label{sec:num}
\subsection{Uniform velocity field (Homogeneous medium)}

We first present a basic example of simulation of wave propagation
in a homogeneous medium to verify RBF-FD and FDM implementations.
Since RBF-FD can mimic FDM when the same grid layout is used, we can
compare the solution obtained with RBF-FD and FDM, to analyse
both methods and also compare the effect of ABCs.

We define the problem on a square domain with dimensions $(\unit[500]{m},
\unit[500]{m})$. The wave velocity is set to $v = \unit[3000]{m/s}$
and is kept constant, implying a constant nodal spacing of
$h = \unit[1.1]{m}$ which gives $N = 248 572$ nodes.
The source of the Ricker's wavelet is
defined to be $(x_s, z_s) = (\unit[150]{m}, \unit[150]{m})$, and the
shape parameter used was $\sigma_R=\unit[0.00147]{s^{-1}}$.

A grid with a comparable number of nodes $N_{\text{FDM}} = 250 000$
(with nodal spacing of $h_{\text{FDM}}=\unit[1]{m}$) was used in FDM simulation.

Stencils for RBF-FD were computed as $n=7$ closest nodes (including the node itself).
and shape parameter for Gaussians is $\sigma_B=\unit[70]{m}$.

We used a small enough time step of
$dt=\unit[0.000098]{s}$ to obtain a stable solution.
The time step was the same in both methods.

To compare the solutions, they were re-interpolated to the same grid using linear interpolation.
The pressure fields and pressure differences shown are in units of $\unit{N/m^2}$.

\begin{figure}[h]
  \centering
  \includegraphics[width=0.8\textwidth]{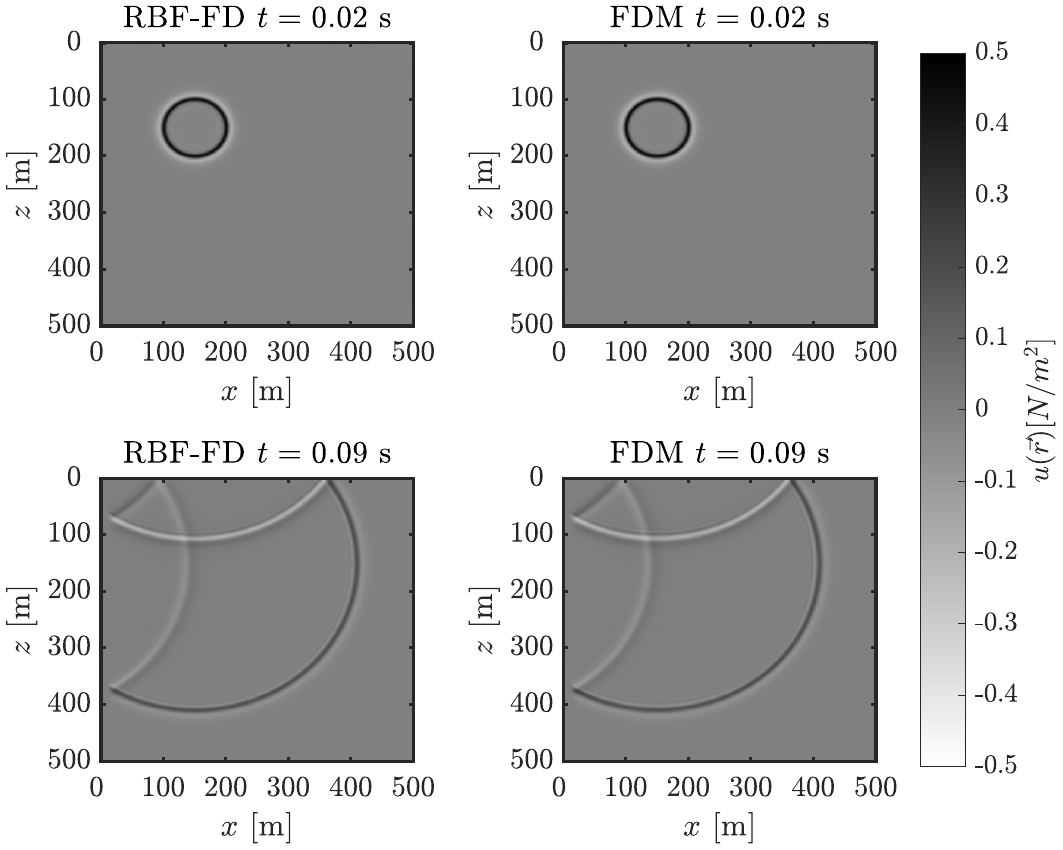}
  \caption{Snapshots of the solution obtained by RBF-FD and FDM method. }
  \label{fig:solution}
\end{figure}

Figure~\ref{fig:solution} shows the wavefield at two different times.
The initial shock propagates in a circular shape until it makes contact with the boundary of
the domain, after which is is completely reflected at the top, but partially absorbed on the left.
The lower two plots illustrate the state after the reflections, where the
effect of absorbing boundary conditions can be observed.

\begin{figure}[ht]
  \centering
  \includegraphics[width=0.8\textwidth]{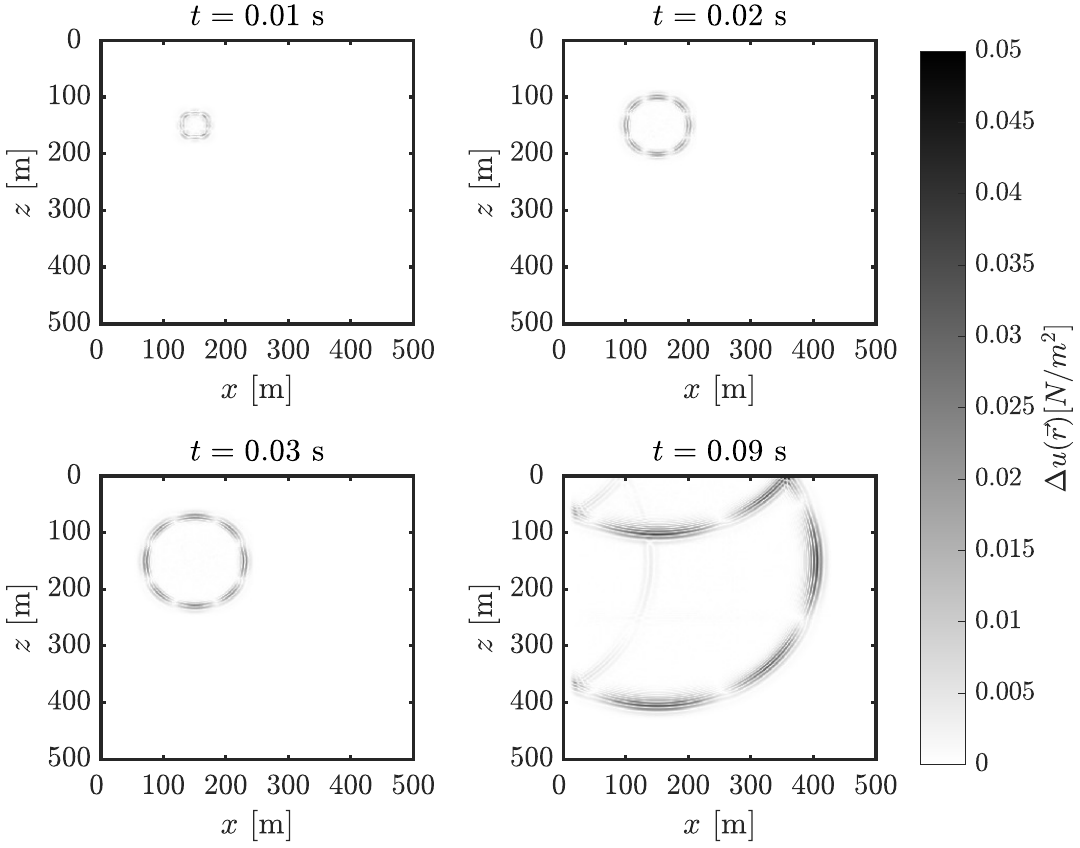}
  \caption{Absolute difference between RBF-FD and 5 point FDM at time four points in time. }
  \label{fig:difference}
\end{figure}

\begin{figure}[h]
  \centering
  \subfloat[Circle in which wavefield is
  interpolated.]{\label{sfig:vel_fixed_C}\includegraphics[width=.44\linewidth]{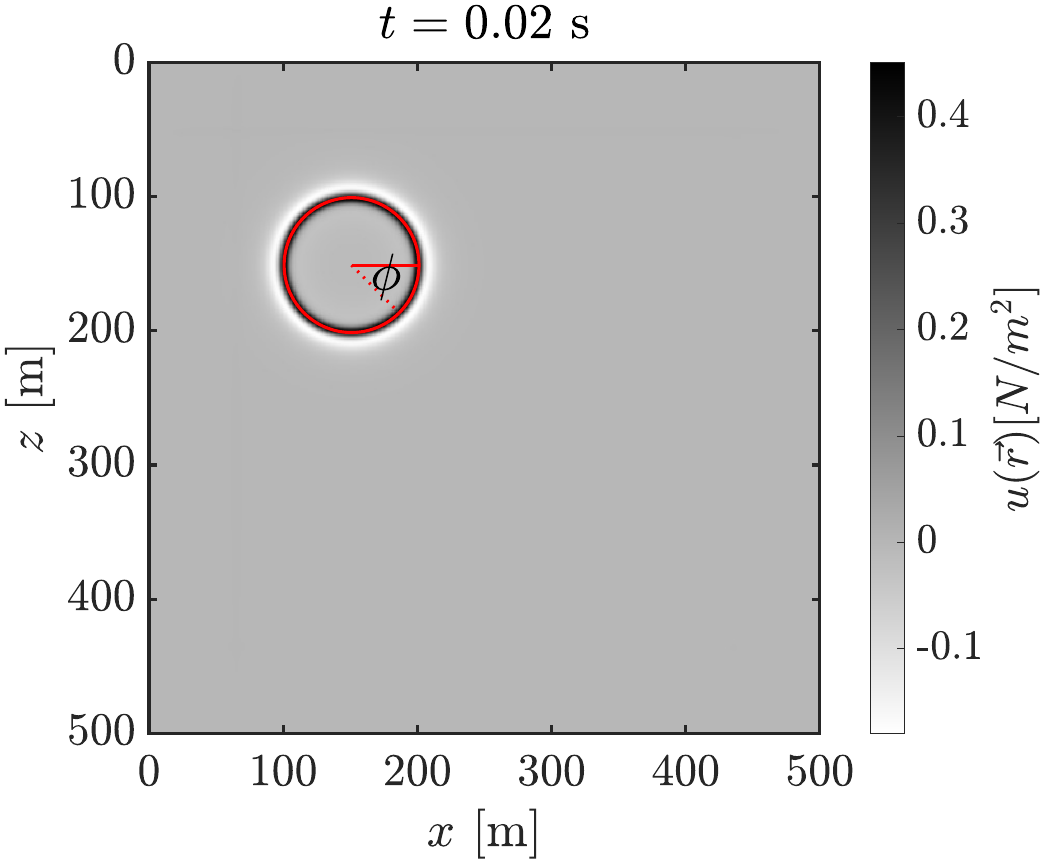}}
   \hspace{0em}%
  \subfloat[Wavefield interpolated in the
  circle.]{\label{sfig:vel_fixed_l}\includegraphics[width=0.45\linewidth]{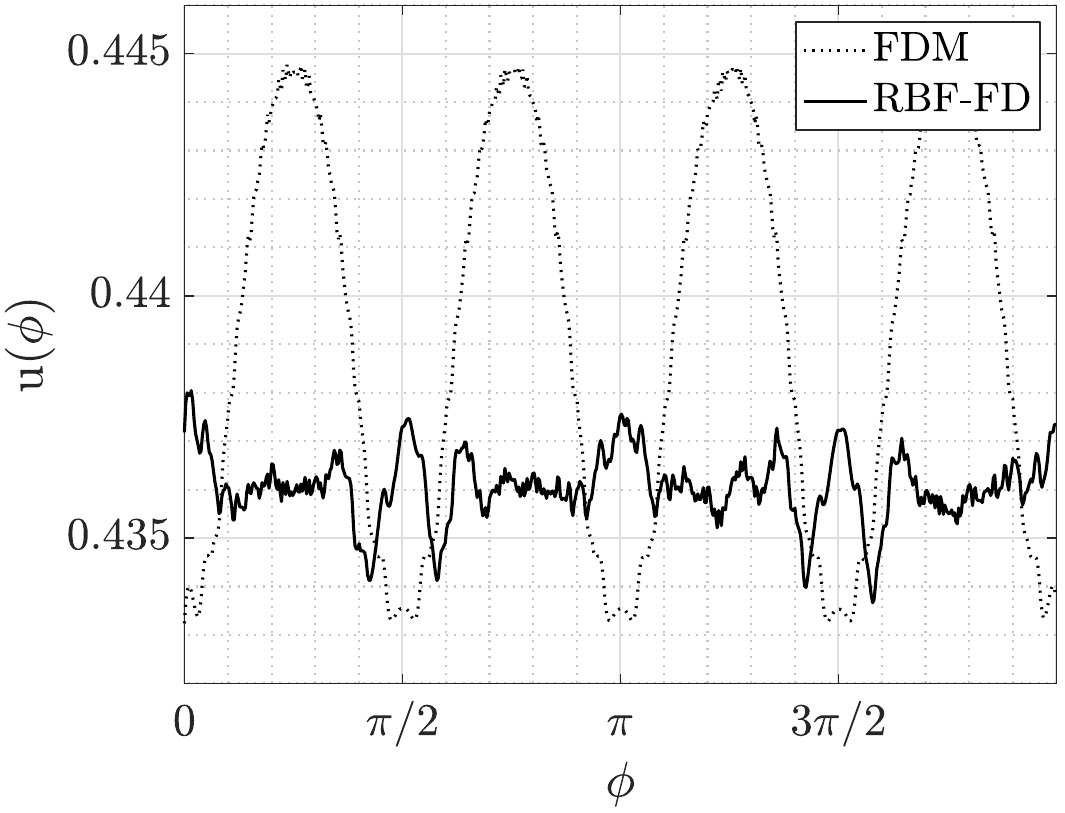}}
  \caption{Symmetry of the RBF-FD and FDM solutions.}
  \label{fig:circle}
\end{figure}

In general, RBF-FD and FDM solution agree well in scope of
error presented in Figure~\ref{fig:difference}. However, in first three plots
of Figure~\ref{fig:difference} one can observe periodic difference between both
solutions on the wave circle. To analyse this phenomenon a plot of the wave
field on the
circle centred at the origin of the source is presented in
Figure~\ref{fig:circle}.  It would be expected that the
displacement fields
are constant on this circumference, as the wave is propagating symmetrically. However, as can be
observed in the right plot in Figure~\ref{fig:circle}, FDM method displays significant
discrepancies from the expected symmetry. While RBF-FD also doesn't provide perfect rotational
symmetry, the discrepancies are noticeably smaller. This difference between methods might be
explained by the larger number of support nodes and  more symmetric placement
employed by
RBF-FD method in comparison to FDM.

\begin{figure}[h]
  \centering
  \includegraphics[width=.7\textwidth]{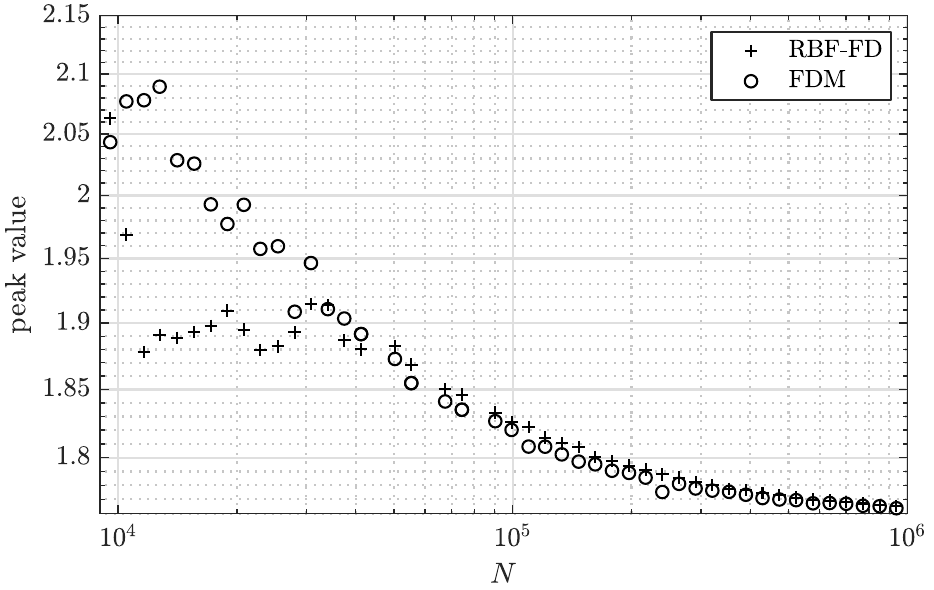}
  \caption{Peak value at $t=\unit[0.03]{s}$ with respect to the number of nodes.}
  \label{fig:convergence}
\end{figure}

The peak value
at time
$t=\unit[70]{ms}$ with respect to the number of nodes is for both methods presented in Figure~\ref{fig:convergence}, where it can be seen that both methods converge to the same value. 

\begin{figure}[ht]
  \centering
  \includegraphics[width=0.7\textwidth]{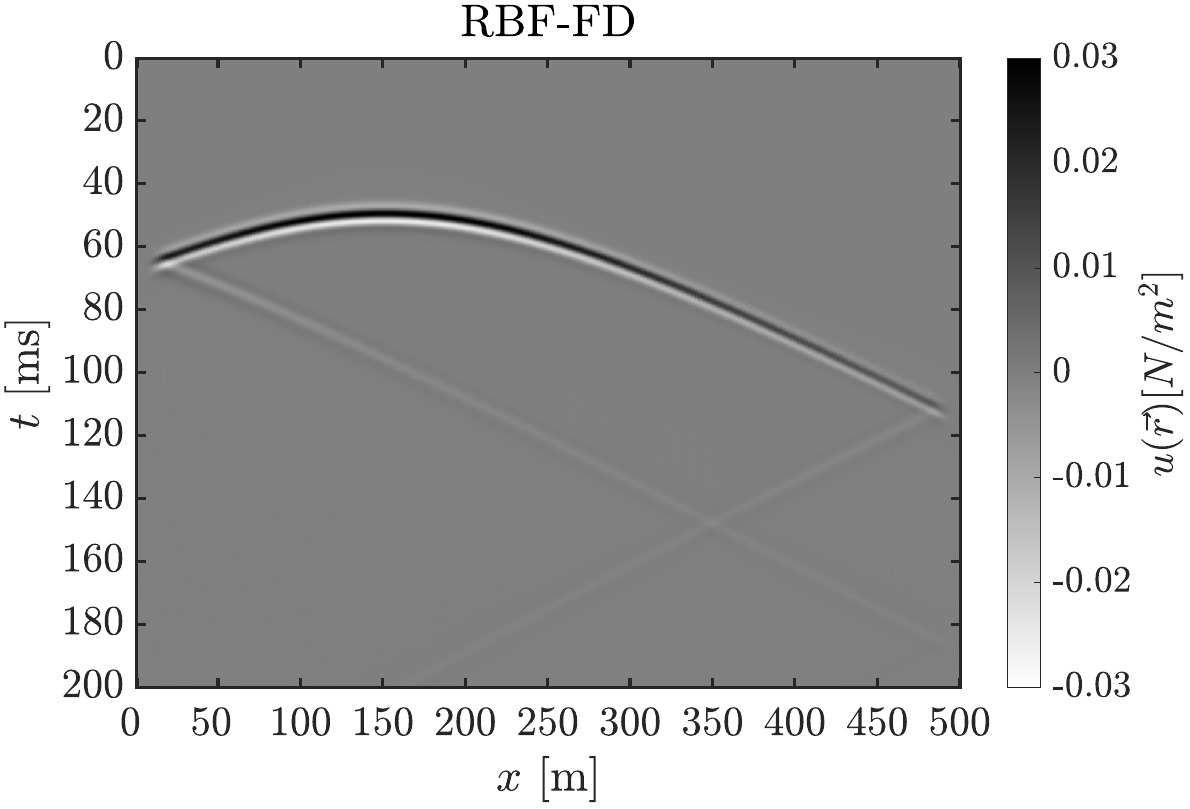}
  \caption{Seismogram obtained with the RBF-FD.}
  \label{fig:seismogram}
\end{figure}

In geophysics there is special significance to the values of the wavefield at the top boundary -
at Earth surface, which is represented with seismogram, i.e. the time
evolution of the wavefield values at the top
boundary. In the
Figure~\ref{fig:seismogram} the $x$ axis corresponds to the horizontal spatial dimension, while
$y$ axis represents the temporal dimension.

In summary, as expected, in this simple case both method produce comparable and
convergent solutions.
% However if for example use a source with higher frequency numerical artifacts would begin to
%appear, as the number of nodes per characteristic length would be too low.

\clearpage

%=====================================================
\subsection{Two-layer velocity model}
In next step we consider a two-layer velocity model. The difference in velocities between layers
suggest different distance between nodes as change of velocity
causes the wavelength  to change.
To evade
numerical artifacts it
is important that an sufficient amount of nodes (10 -- 20) is present per
wavelength~\cite{geiger, alford1974accuracy}.
Using the RBF-FD meshless method,
there aren't any restrictions on node placement, which gives it an advantage over conventional
methods. Consequently variable node density in relationship to the velocity field is easily
implemented.

\begin{figure}[h]
  \centering
  \subfloat{\includegraphics[height=6.5cm]{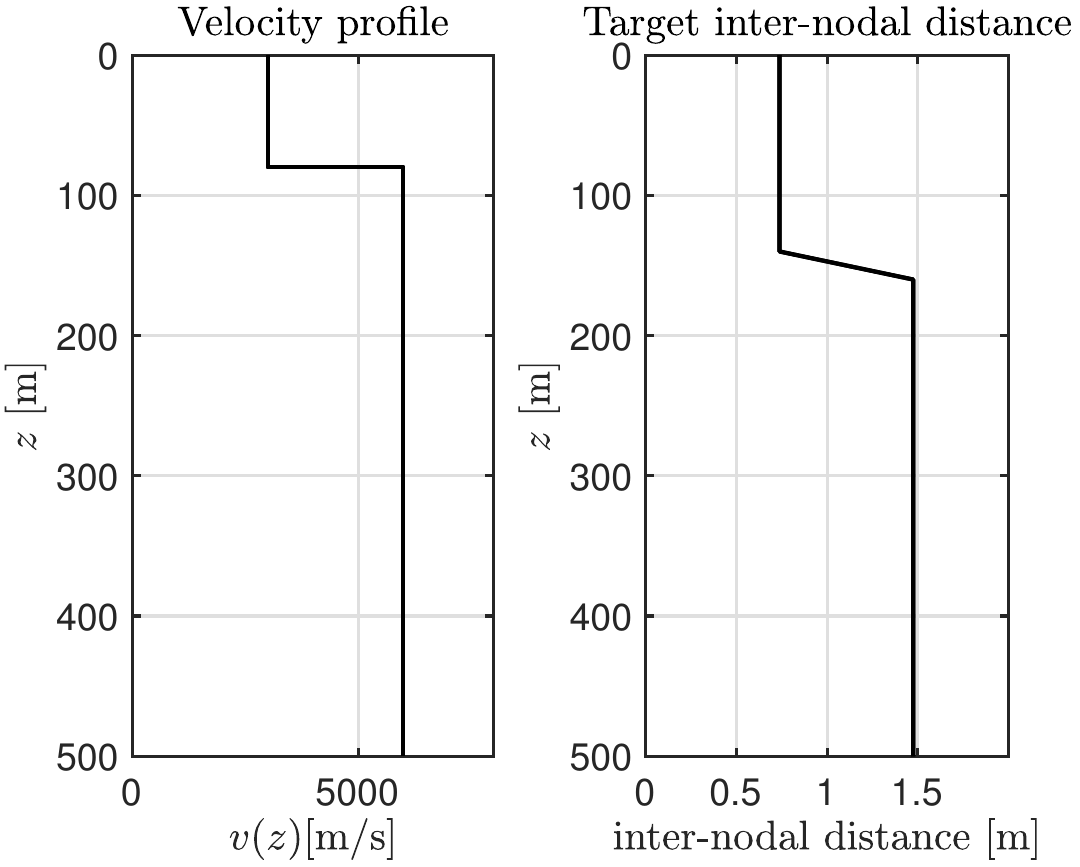}} \hspace{2em}%
  \subfloat{\includegraphics[height=6.5cm]{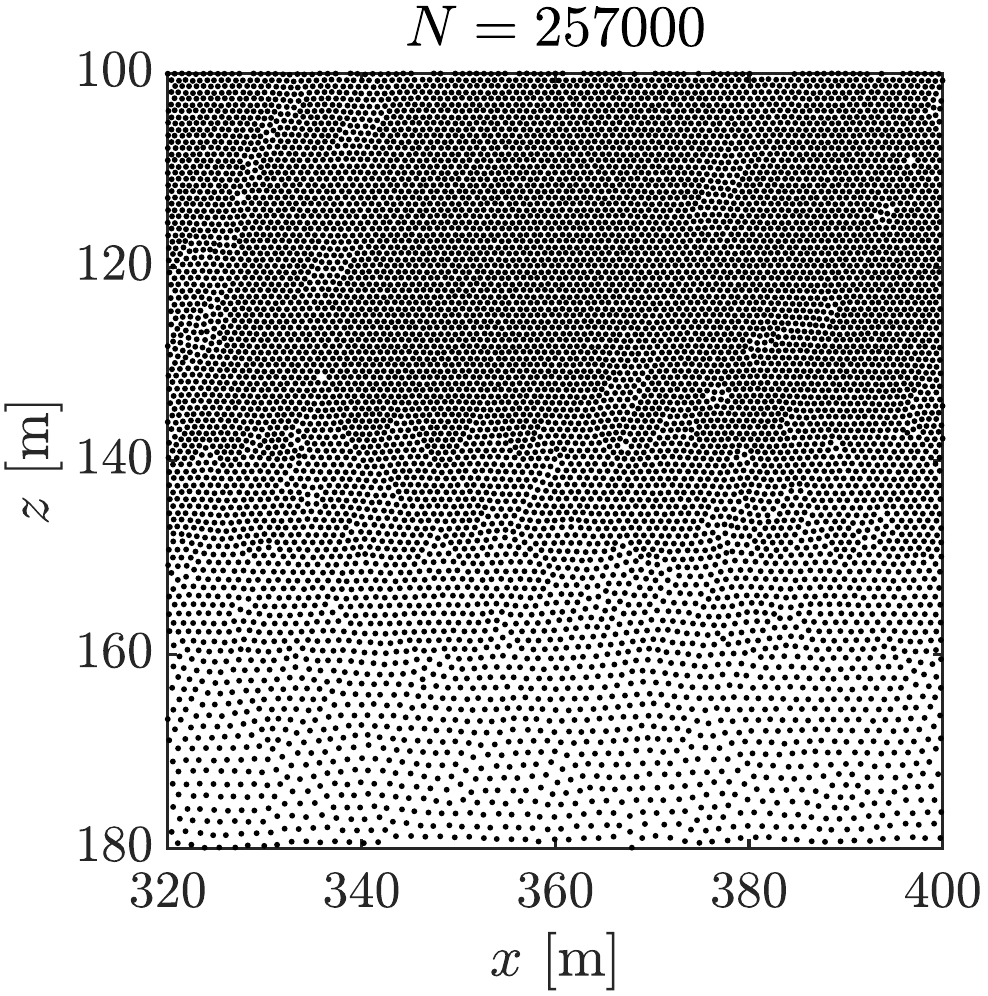}}

  \caption{$z$ cross-section velocity profile (left), $z$ cross-section inter-nodal distance function (center) and snapshot of node placement (right).}
  \label{fig:case2_domain}
\end{figure}

In Figure~\ref{fig:case2_domain} the $z$ cross-section velocity profile and
corresponding RBF-FD
nodes are presented. The jump in velocity happens at depth of $\unit[80]{m}$. It can be observed that the jump in inter-nodal distances doesn't
directly follow the jump
in velocity. The jump happens at depth of $\unit[150]{m}$. The inter-nodal distance function $h(z)$
is made continuous by application of a moving average over the step function
\begin{equation}
	h(z)= \text{moving average}( \unit[0.737843]{m} + H (z-\unit[150]{m})\text{ }\unit[0.737847]{m}),
\end{equation}
where $H$ is the Heaviside step function.
 The displacement of the jump in node density is a necessary compromise which will be discussed in more detail with the
presentation of the results.
Dimensions of the domain are  $(\unit[500]{m},
\unit[500]{m})$. For RBF-FD method time
step is set to $dt = \unit[0.000058]{s}$ and for FDM method it is set to $dt = \unit[0.000167]{s}$.
The source is located at $(x_s, z_s) = (\unit[250]{m},
\unit[200]{m})$, and parameter
$\sigma_R=\unit[0.00106]{s^{-1}}$ was used for the
Ricker's wavelet. As stated previously parameter $\epsilon=\unit[4]{m}$ was used.
%\begin{figure}[ht]
%\centering
%\includegraphics[scale=0.25]{Figures/case2/domain_.png}
%\caption{Velocity profile (left) and node placement (right) \cmj{Scale on the right side is
%fliped.}.}
%\label{fig:case2_domain}
%\end{figure}

\begin{figure}[ht]
  \centering
  \includegraphics[width=0.75\textwidth]{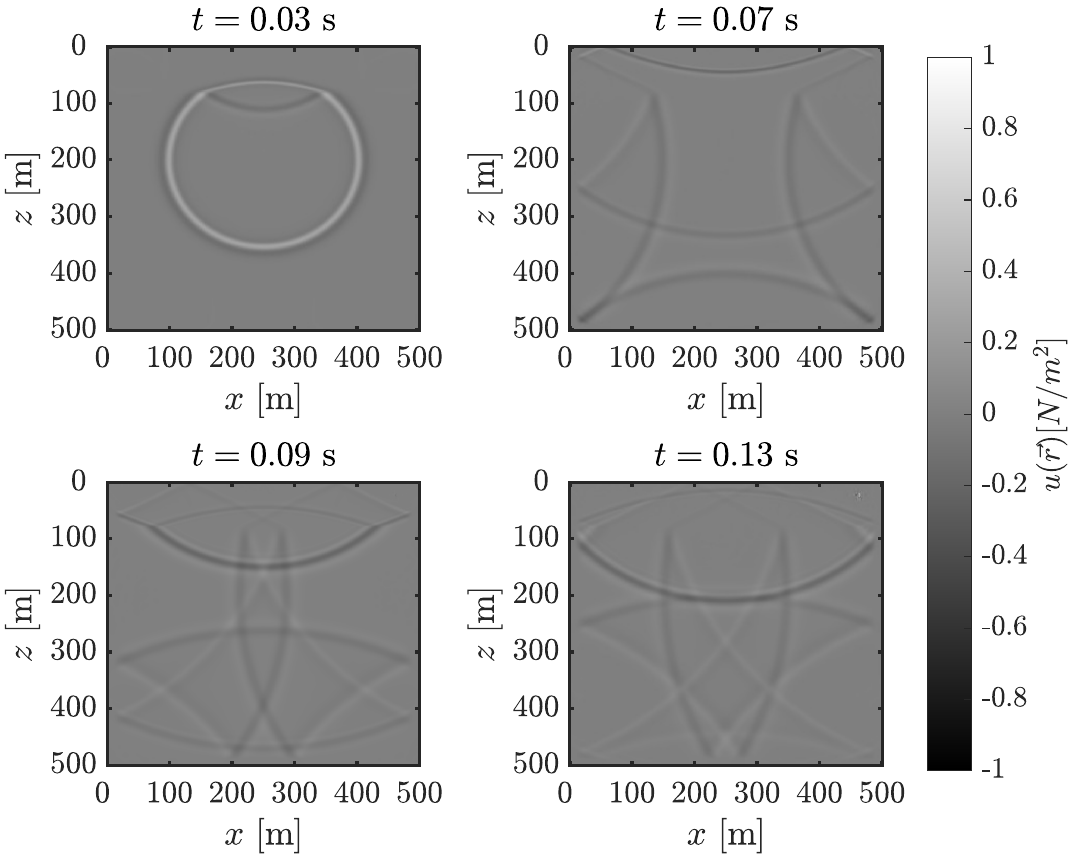}
  \caption{RBF-FD solution snapshots.}
  \label{fig:case2_solution}
\end{figure}

Snapshots of the wavefield are presented at 4 different times in Figure~\ref{fig:case2_solution}.
Again we observe the reduced reflections from the boundary.  A new phenomenon present in this
case is the partial reflection at $y=\unit[250]{m}$.  This is most clearly visible at time
$t=\unit[0.03]{s}$, where this is the only reflection resent in addition to the
original wave
propagating from the source. Decreasing of wavelength can be observed as well.
%\cm{from here on I cannot follow the logic of results presentation. Ok, in above discussion we
%demonstrate the solution of the case, that is fine. Then we start to discuss difference between
%the FDM and RBF and different epsilons. In Figure 9 we have RBF-FDM for case eps = 8, why do we
%need this figure? What does it tell us? In fig 10 we have seismogram ... basically continuation
%of presentation of results, that is ok, but it should be next to figure 8. Then we have Fig 11,
%x
%cross section ... where we clearly demonstrate that FDM is stable and RBF-FD is not. And finally
%we have fig 12, where we show the ripples in FDM solution. Well I think it is a strange mix of
%eps 8 and eps 1.5 results without a clear idea what we want to sell to the reader. What if we
%first show eps 1.5 and 8 RBF solution as 2xfig 8 + 2xfig 10. And the we start discussion about
%difference between FDM and RBF, and support it with, let's say 2xFig 11 together with
%2xfig:cross
%section x = 250, y, where we could, hopefully, observe the FDM ripples. Finally we have to
%adequately comment this phenomena. -- before you rush into matlab we should discuss that}
%Figure~\ref{fig:case2_dif} displays the absolute difference between methods
%
%Figure~\ref{fig:case_2_seismogram} displays the seismogram for this case.
%\begin{figure}[ht]
%\centering
%\includegraphics[width=0.8\textwidth]{Figures/case2/dif1.jpg}
%\caption{Difference between FDM and MM. $\epsilon= 8.0$. }
%\label{fig:case2_dif}
%\end{figure}
%\begin{figure}[ht]
%\centering
%\includegraphics[width=0.8\textwidth]{Figures/case2/seiz_MM1.jpg}
%\caption{seismogram meshless method.$\epsilon= 8.0$.}
%\label{fig:case_2_seismogram}
%\end{figure}

%\begin{figure}[ht]
%\centering
%\includegraphics[width=0.8\textwidth]{Figures/case2/cross1.jpg}
%\caption{Cross-section $y=200$ meshless method.$\epsilon= 8.0$.}
%\label{fig:case_2_cross}
%\end{figure}

%\cm{Perhaps we could add simple analysis of FDM vs RBF-FD performance with respect to the
%$\epsilon$ - but in case 1 -- all basic tests belong in case 1}

\begin{figure}[ht]
  \centering
  \includegraphics[width=0.8\textwidth]{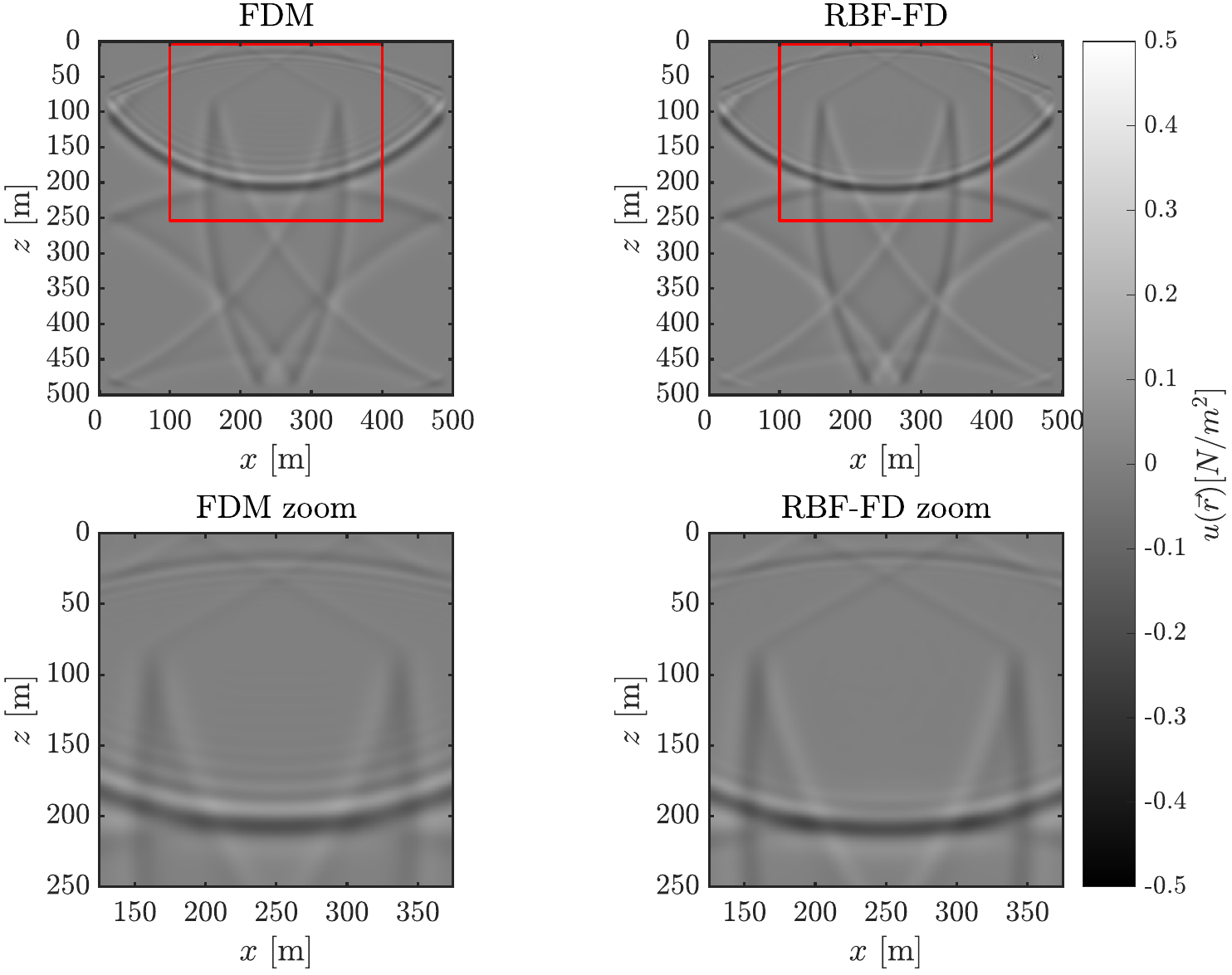}
  \caption{Snapshot at time 0.13s.}
  \label{fig:case2_prob}
\end{figure}

The results from RBF-FD are compared to those from FDM in Figure~\ref{fig:case2_prob}. Both
methods are tested on discretization with approximately $250 000$ nodes, however RBF-FD method
distributes nodes as described at the beginning of this subsection in contrast to homogeneous grid used by FDM.

In Figure~\ref{fig:case2_prob} artificial ripples are present on snapshot of FDM solution, which
do not develop when the RBF-FD solution is employed.

Such errors are caused by insufficient node density. Using RBF-FD this problem was avoided, by
increasing the density in upper region while simultaneously decreasing the density in lower
region, which does not reduce the accuracy as much since the wavelength in the lower
region is larger.

As stated before, the step in velocity and the decrease in node density do not align exactly. This
necessitated by the fact that sudden jumps in nodal density cause new numerical errors and the
fact that nodal density must be sufficiently high everywhere on the domain. If just a moving
average of the velocity field would be used as the basis for target inter-node distance function, the second condition would not be met in narrow
region on top of the point of velocity step.
For best results it proved necessary to delay the jump in node density in comparison to point of jump
in velocity field.

In addition to slightly increasing the amount of nodes necessary, another downside is introduced.
If the node density used with RBF-FD directly followed the velocity field, the solution would
actually be more stable than one provided by FDM. This can be understood by looking at the
stability criterion for FDM:

\begin{equation} \label{crit}
dt\varpropto\frac{dx}{v}.
\end{equation}
When using FDM the only change in comparison to case with constant velocity is the increase of
the velocity in the lower half of the domain. Flowing the criterion, this results in smaller
required time-step for all of the simulation. When RBF-FD is used depending on nodal distribution
two cases are possible:
\begin{itemize}
  \item If the nodal density follows the velocity field directly, meaning $dx\propto v$ , the
  velocity dependance of the criterion cancels out. This means that areas of high velocity do not
  dictate the use of shorter time steps in simulation.
  \item In case where the density field does not follow the velocity field directly, we lose the
  stability advantage. In the narrow area below the point of velocity step,  the node density is unchanged while the
  velocity increases, this results in same necessity for decrease in time step. The time
  step actually needs to be even smaller than one required by FDM, as $dx$ in the dense region is
  smaller than one used by FDM, which reduces time step further as $dt\propto dx$ follows from
  stability criterion.
\end{itemize}
% \begin{figure}[ht]
% \centering
% \includegraphics[width=\textwidth]{Figures/probl_seiz.jpg}
% \caption{Seizmogram.}
% \label{fig:case2_seiz}
% \end{figure}
All of the discussed assumes the stability criterion for FDM is at least to a factor also valid
for RBF-FD. The need for a lower time-step might be a cause for concern, however if one would
want similar performance to one achieved using RBF-FD, FDM with grid of higher density would be
necessary. Not only would this drastically increase the number of computational nodes, the
time-step would also have to match the smaller one used by RBF-FD, as the $dx$ in criterion would
now be the same for both methods.

\begin{figure}[ht]
	\centering
	\includegraphics[width=0.9\textwidth]{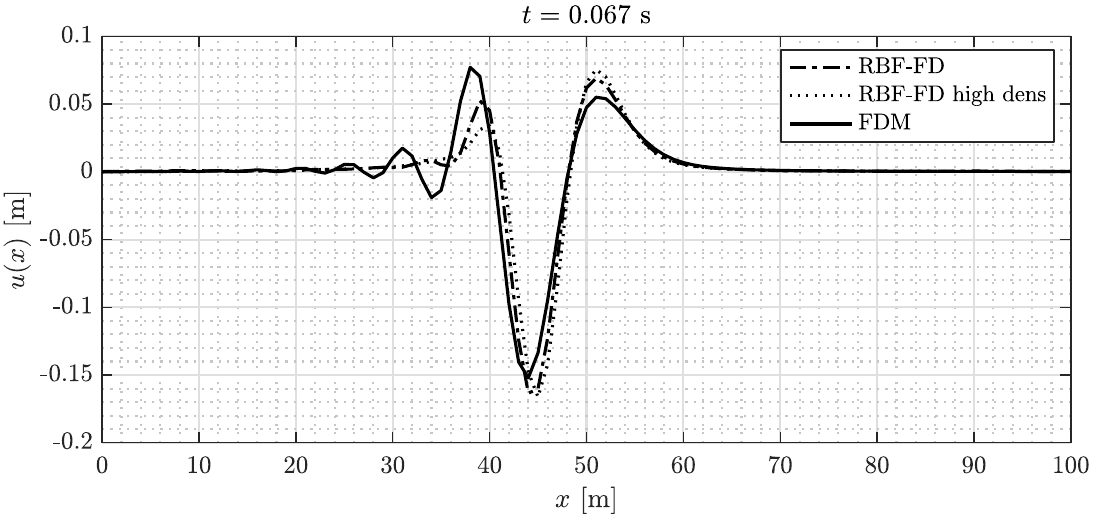}
	\caption{Cross-section.}
	\label{fig:case2_cross}
\end{figure}

While the difference was already very clear in Figure~\ref{fig:case2_prob}, cross-section view
provides even more detailed picture. We look at the cross-section at $x=\unit[250]{m}$ in
Figure~\ref{fig:case2_cross}. Snapshot is provided at time $t=\unit[67]{ms}$.

\begin{figure}[h]
	\centering
	\subfloat{\label{sfig:vel_fixed_C2}\includegraphics[width=.5\textwidth]{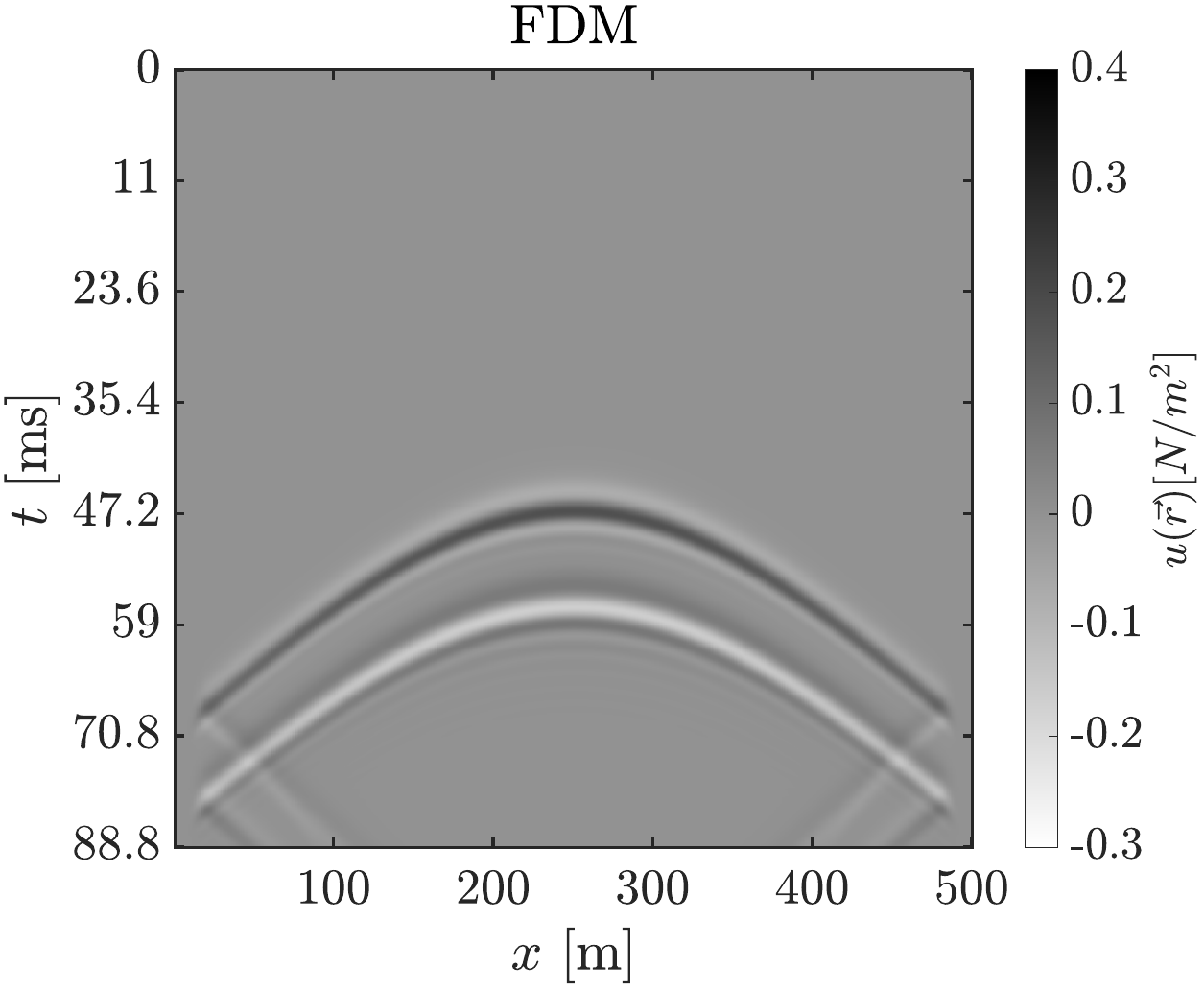}}
	\hspace{0em}%
	\subfloat{\label{sfig:vel_fixed_l2}\includegraphics[width=.5\textwidth]{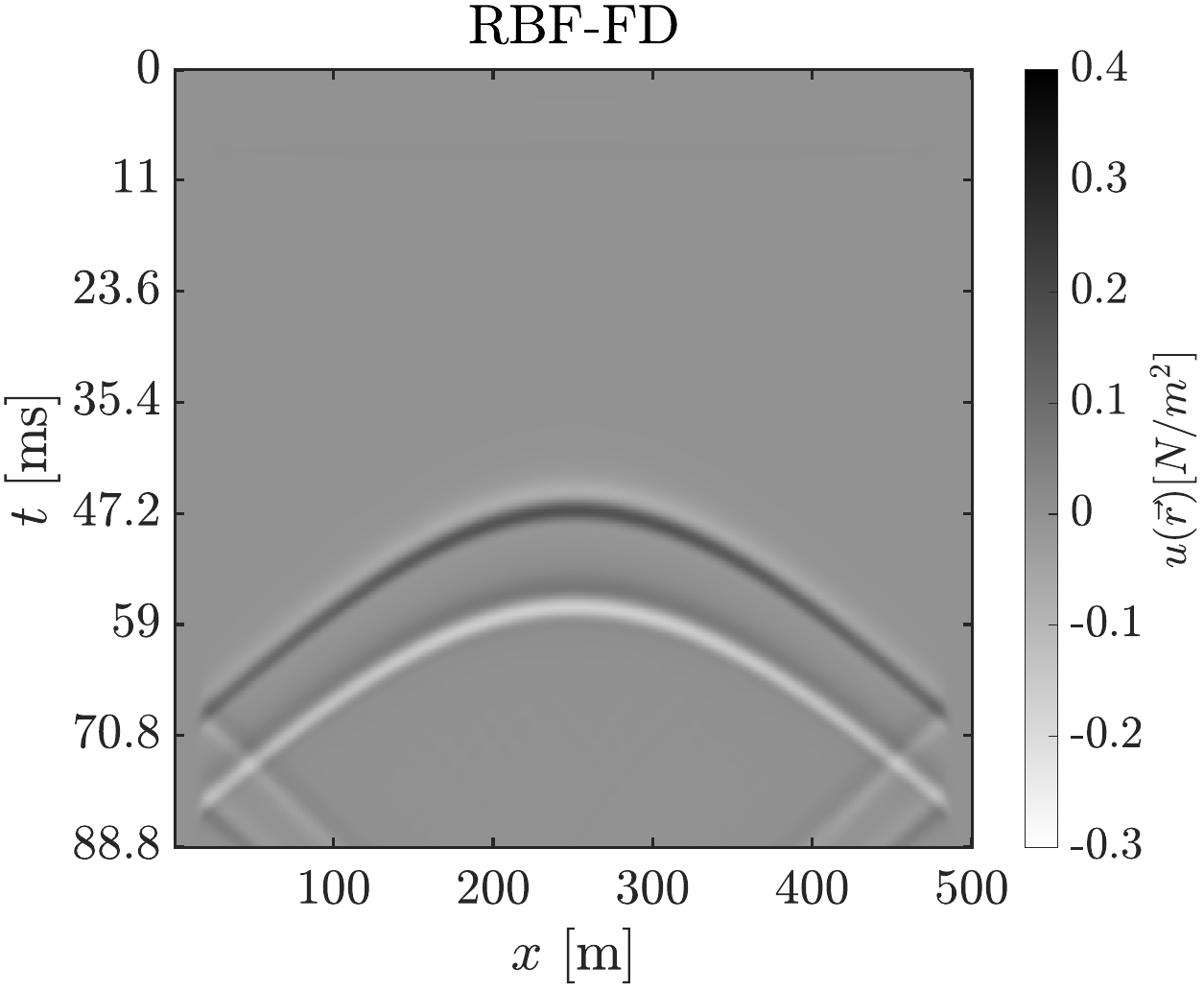}}

	\caption{Seismogram two-layer model.}
	\label{fig:2_seis}
\end{figure}

The RBF-FD solution displayed here always provided at least 11 nodes per
characteristic
distance of the wave. For reference Another RBF-FD solution is added, where the density is
higher, namely at least 15 nodes per characteristic distance.
%Here the ripples discussed before can be clearly observed on FDM solution.

To conclude the analysis of this numerical example, seismograms are provided for both methods in
Figure~\ref{fig:2_seis}. Again, we can make similar observations about improved
accuracy of RBF-FD
method in this case.
%
%To Summarize, RBF-FD method provides an elegant solution to problems with numerical artifacts,
%that arise with conventional methods in cases where required node density changes dramatically
%with location, without much extra work on the side of the researcher.
% Another detail that has to be addressed is the difference in solutions on the right-most wave
%front on un-magnified cross-section at time $t=0.067$. The difference is quite substantial, and
%one might assume FDM solution is there more accurate as it is in the area where FDM has much
%greater node density, as because of higher velocity RBF-FD reduces its node density there.
%However the inclusion of the RBF-FD solution with higher density contradicts this logic, as the
%solution moves even more away from FDM one.
%With all possible factors considered we have concluded, that the difference is caused by the
%larger time-step used by the FDM solution. In cases of high velocity the time step becomes of
%increased importance to the accuracy of the method. This was in more detailed while discussing
%the first numrical example and is demonstrated on Figures~\ref{fig:VEL++FDM}
%and~\ref{fig:VEL++MM}.

%=====================================================
\subsection{Marmousi velocity model}
\begin{figure}[h]
  \centering
  \subfloat{\includegraphics[width=.58\linewidth]{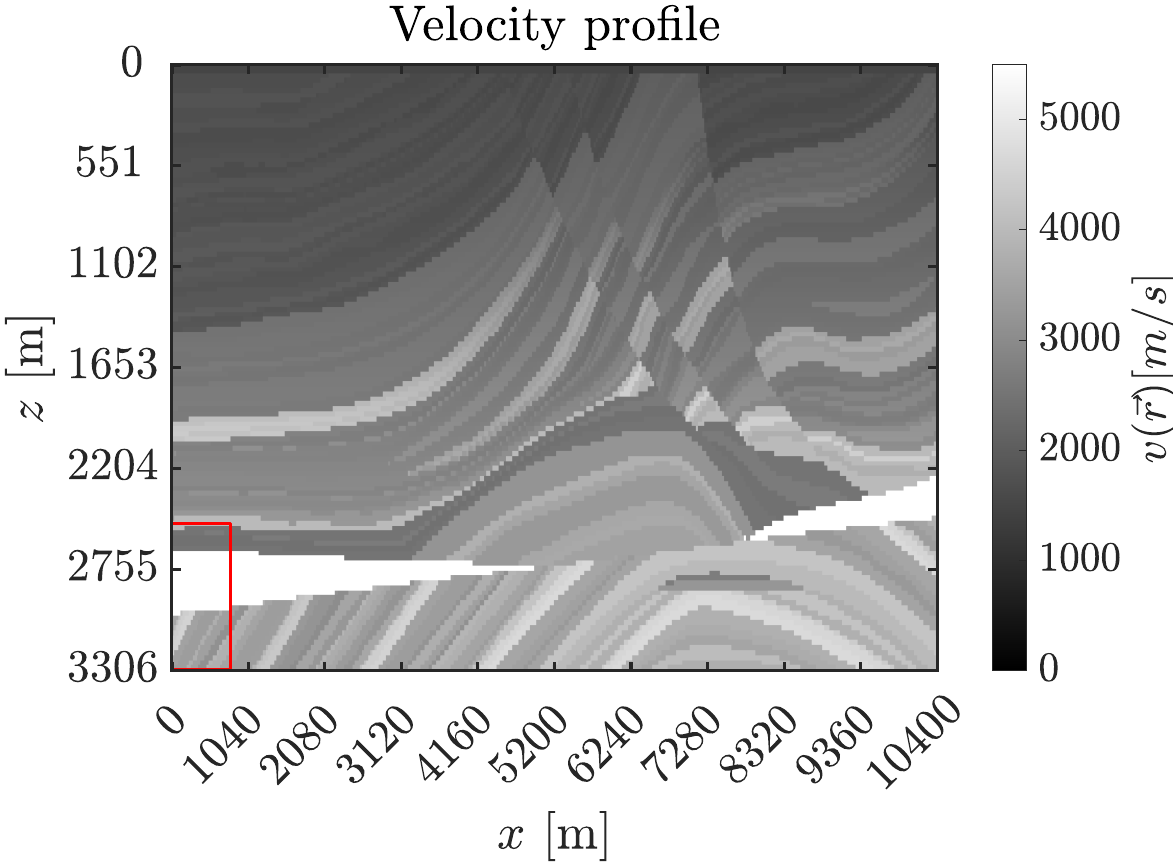}} \hspace{0em}%
  \subfloat{\includegraphics[width=.42\linewidth]{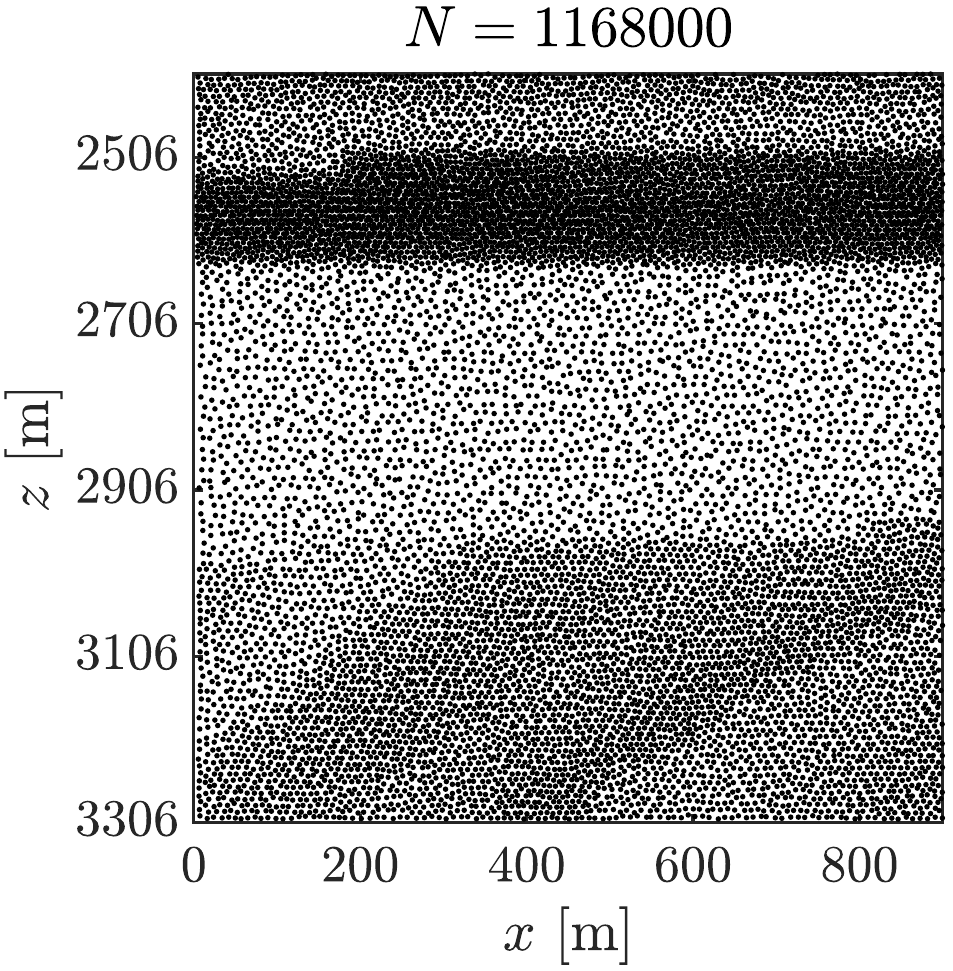}}

  \caption{Velocity profile (left) and node placement (right).}
  \label{fig:case3_domain}
\end{figure}

\noindent For the last numerical test we look at a more complicated example of Marmousi velocity
model~\cite{Versteeg1994}, displayed in the left of Figure~\ref{fig:case3_domain}. Similarly as in section
on Numerical test 2, the node density can be related linearly to the velocity field, however in this case without displacing and smoothing, as we will ensure enough nodes in high-velocity area for stable simulation. Since the
data points of the velocity model do not generally align with the positions of computational
nodes, Sheppard's scattered data interpolation is used to determine the density at the required
positions.

\begin{figure}[ht]
  \centering
  \includegraphics[width=\linewidth]{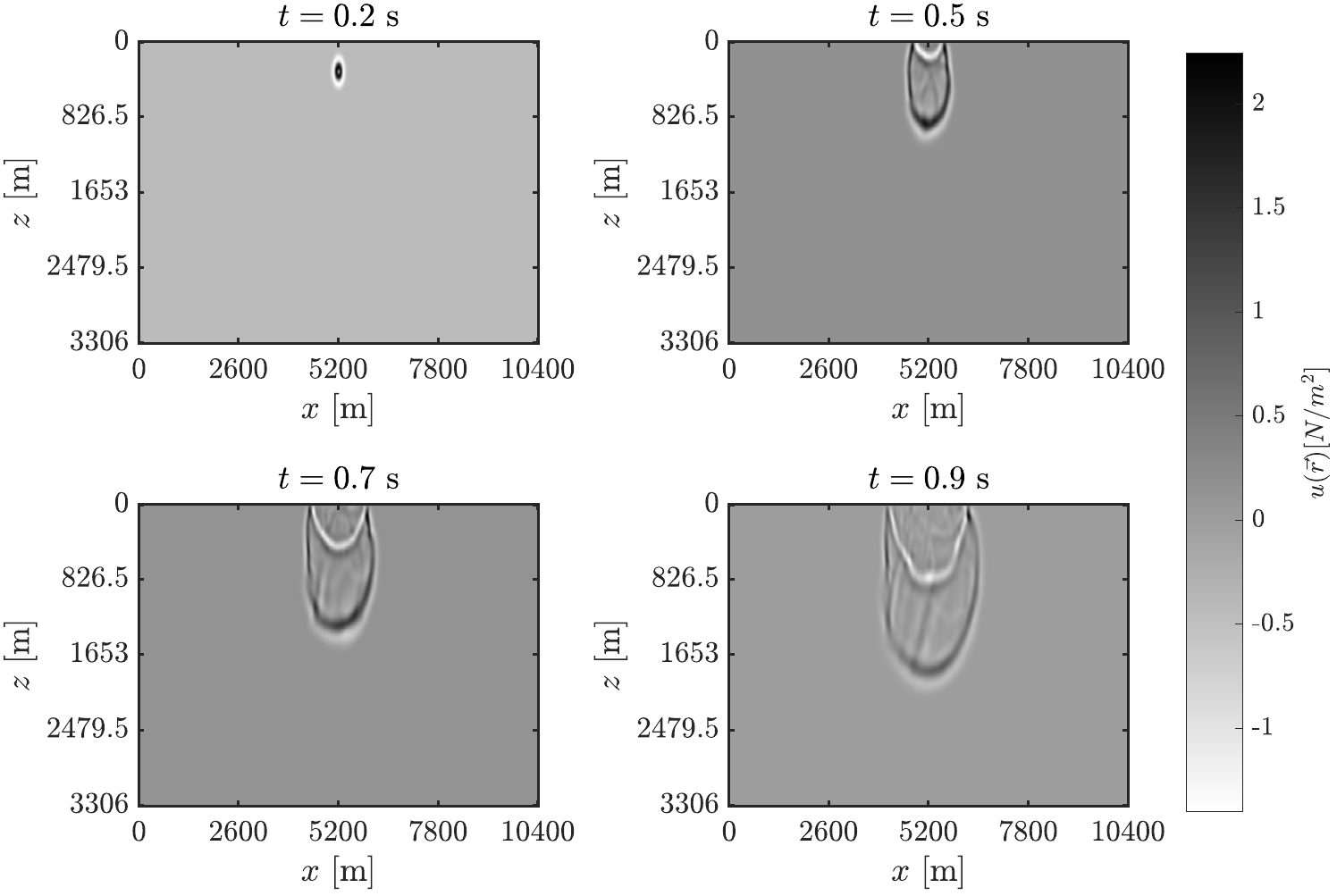}
  \caption{RBF-FD solution snapshots for the Marmousi velocity model.  }
  \label{fig:case3_MM}
\end{figure}

On the right side of Figure~\ref{fig:case3_domain} a zoomed view of the node placement is
displayed. This section is marked with red rectangle on the velocity model. In this test the size
of the domain is $(\unit[10400]{m}, \unit[3306]{m})$, the time step is set to
$dt=\unit[0.00087]{s}$ and the source is positioned at $(\unit[5200]{m}, \unit[330.6]{m})$
The wavefield snap-shot at different time-intervals have been shown in Figure
\ref{fig:case3_MM}.
We can observe the distortion of the primary wave and its reflections caused by the velocity
field. The solution is also free of any obvious numerical artifacts.

\begin{figure}[ht]
  \centering
  \includegraphics[width=0.9\textwidth]{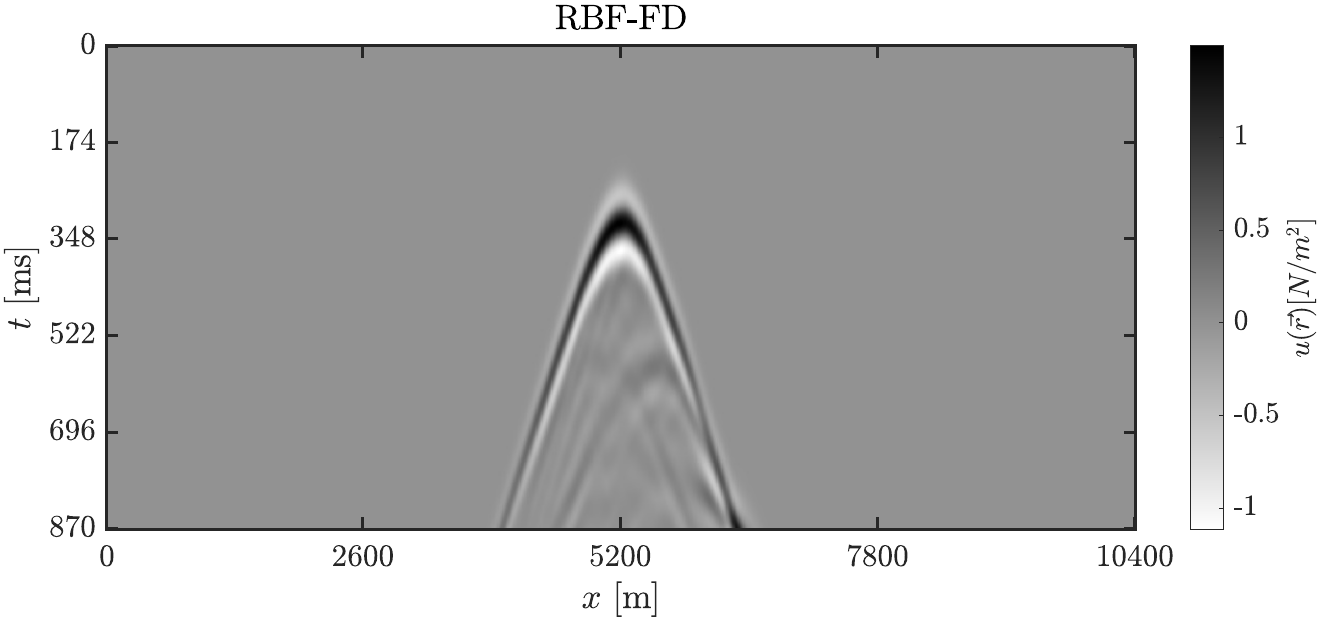}
  \caption{Marmousi seismogram.}
  \label{fig:case3_seis}
\end{figure}

We provide the seismogram for this example in Figure~\ref{fig:case3_seis}. We can again observe
the secondary waves caused by subsurface reflations.

In conjunction with results from previous two cases we can conclude RBF-FD is a viable
alternative to conventional methods, such as FDM. It can be applied to cases with arbitrarily
complex velocity fields and can reduce numerical artifacts without drastically
increasing computational intensity.

\subsection{Irregular domain shape: Canadian Foothill}
\begin{figure}[ht]
	\centering
	\includegraphics[width=0.7\textwidth]{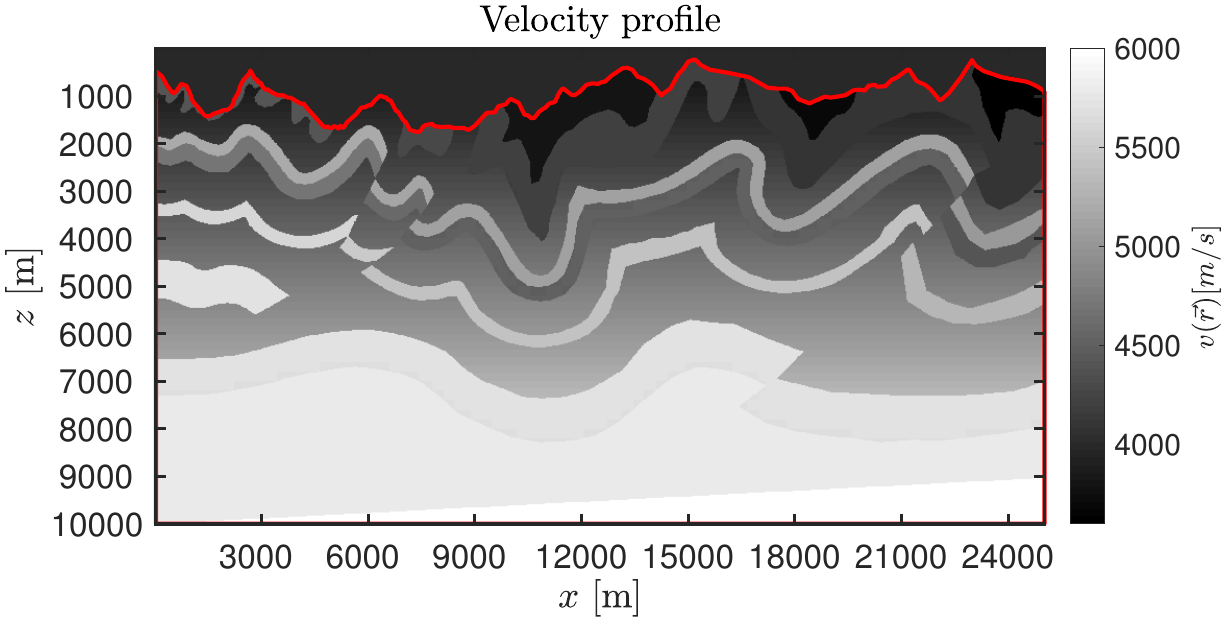}
	\caption{Velocity profile for Canadian Foothill case. Shape of the domain is marked with red line.}
	\label{fig:case4_velocity_profile}
\end{figure}

\begin{figure}[ht]
	\centering
	\includegraphics[width=0.7\textwidth]{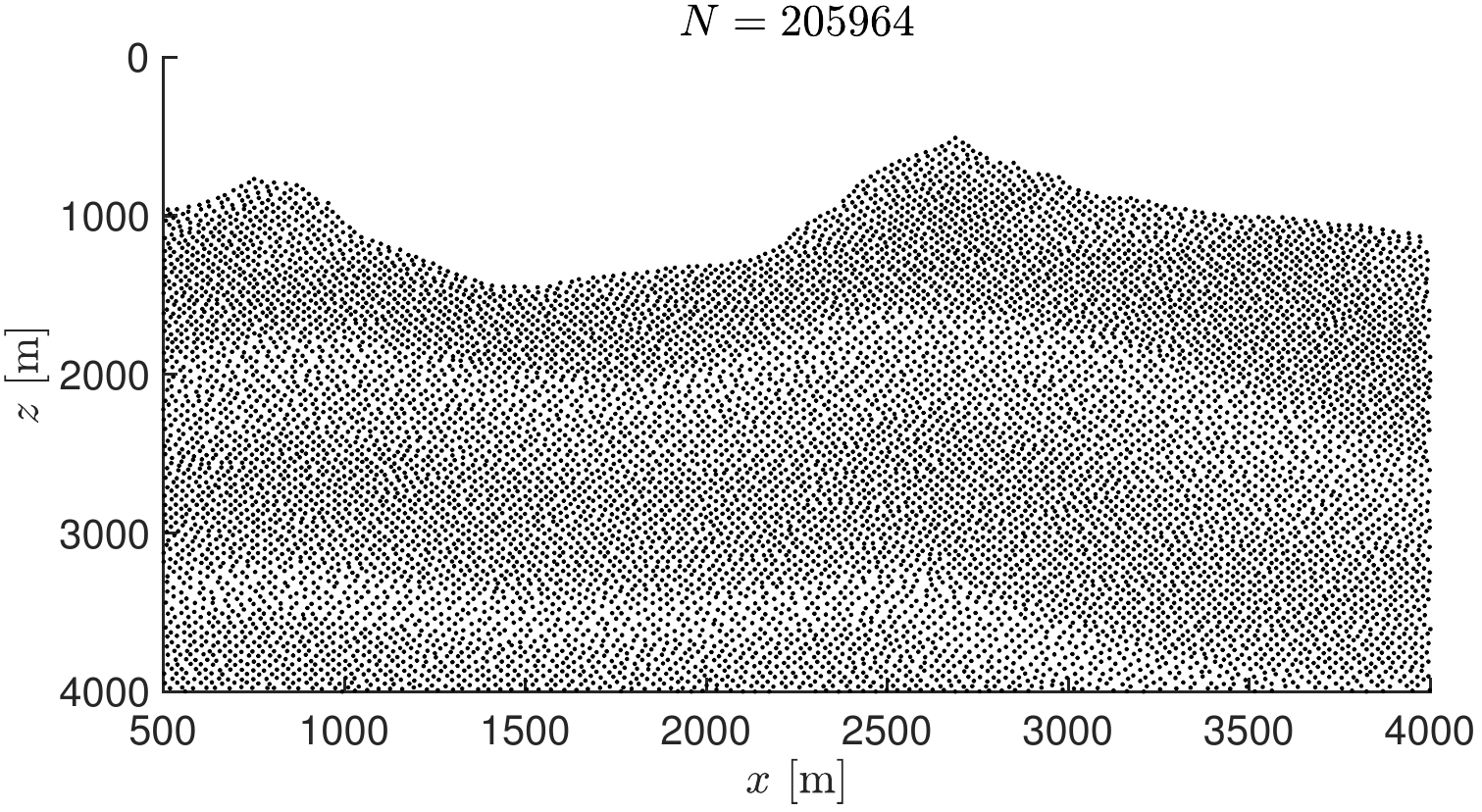}
	\caption{Distribution of nodes, for better visibility only a upper left of the domain is displayed.}
	\label{fig:case4_nodes}
\end{figure}

\begin{figure}[ht]
	\centering
	\includegraphics[width=\textwidth]{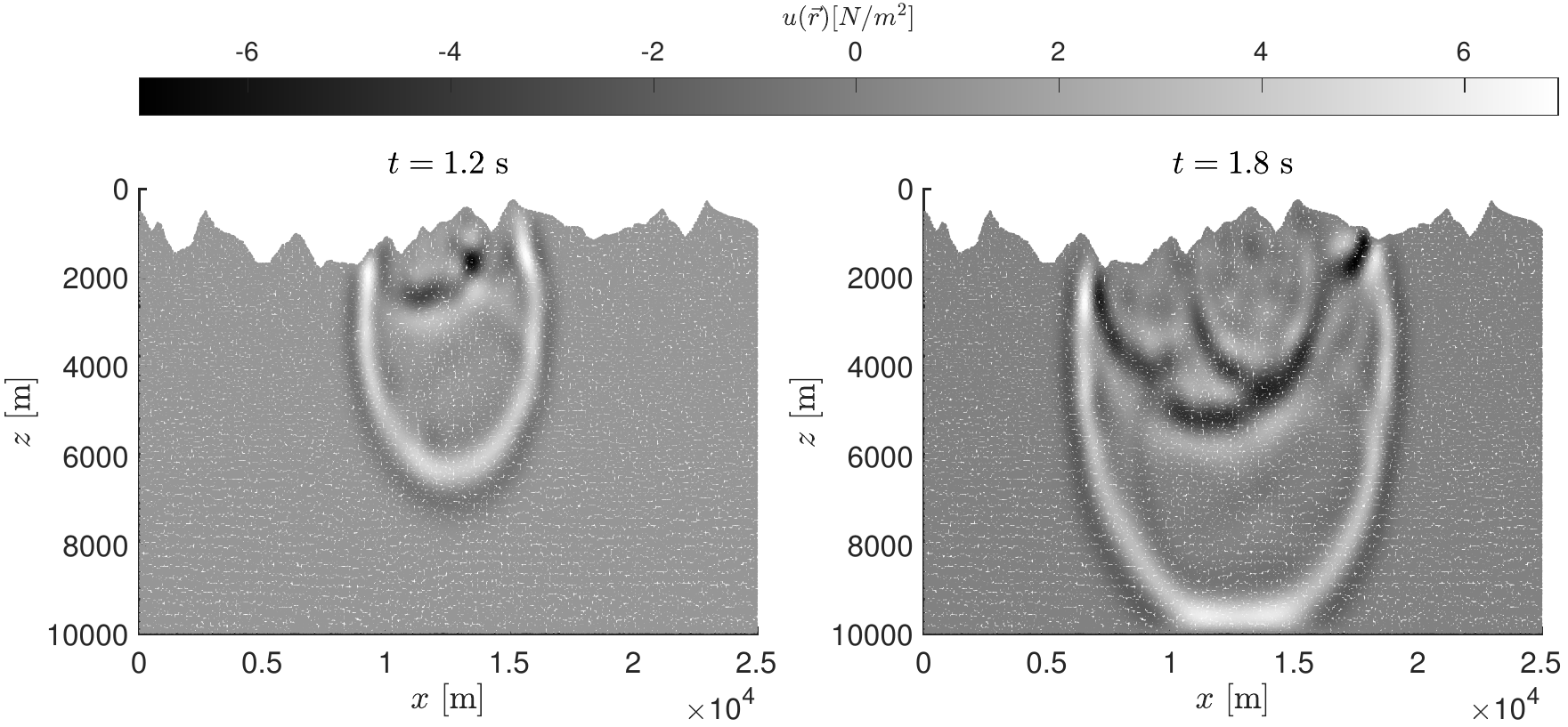}
\caption{RBF-FD solution snapshot for Canadian Foothill velocity profile.}
	\label{fig:case4_snapshot_with_profile}
\end{figure}

To demonstrate performance of presented RBF-FD based solution procedure on irregular domains, the Canadian Foothill model is addressed~\cite{gray1995migration}. The domain is enclosed inside a rectangle of dimensions $\unit[25000]{m}$ times $\unit[10000]{m}$ as presented on Figure~\ref{fig:case4_velocity_profile}. Sea level is at the $\unit[2000]{m}$ mark. The original velocity profile of the model has resolution of $1668$ times $1000$ with data points spaced at $\unit[15]{m}$ horizontally and $\unit[10]{m}$ vertically. 
The case was set up with a point like source at locations ($\unit[12500]{m}$, $\unit[2500]{m}$). RBF-FD method was used with time step $dt=\unit[0.001]{s}$ on $205964$ scattered nodes positioned with~\cite{slak2019generation}(\ref{fig:case4_nodes}). 
First, we solved a simplified model with constant velocity profile $v=\unit[4500]{m}$ that is presented in Figure~\ref{fig:case4_snapshot_with_profile}. In next step, we included also Canadian Foothill velocity profile, results are presented in Figure~\ref{fig:case4_snapshot_with_profile}.

\begin{figure}[ht]
	\centering
	\includegraphics[width=\textwidth]{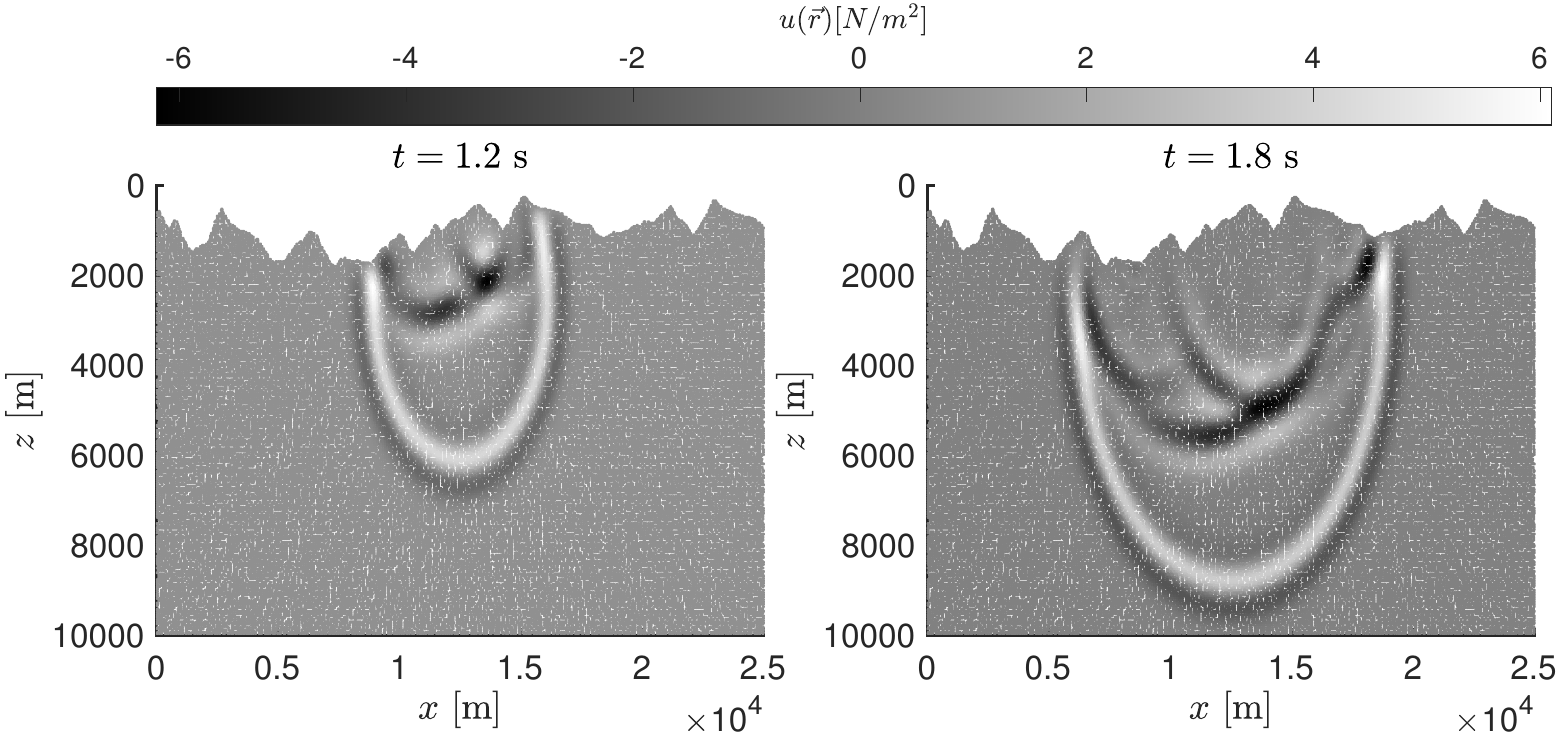}
	\caption{RBF-FD solution snapshot for constant velocity  profile $v=\unit[4500]{\frac{m}{s}}$.}
	\label{fig:case4_snapshot_no_profile}
\end{figure}

%=====================================================
\newpage
\section{Conclusions}
We have investigated a local strong-form meshless method RBF-FD for numerical solution
of 2D time-domain acoustic wave equation in heterogeneous media. The numerical tests performed
here have twofold importance: (a) It is one more-step towards the robustness of the current
understanding of the RBF-FD by exploring the acoustic wave propagation problem, and (b) the
RBF-FD has the potential of being used in large-scale seismic modeling and inversion
applications. Followings are some conclusions we draw from the present study: 

\begin{enumerate}
  \item RBF-FD has the advantage of working with node-distribution, which are adaptive to
  the given velocity-variations. This is a clear advantage over conventional finite difference
  method. Moreover, RBF-FD save the effort through bypassing the steps of mesh-generation and
  preserving its shape trough out the time-iteration, which is an advantage over finite-element
  type methods.
  \item Since the stability-criterion in the RBF-FD method
  can also be adaptive to the velocity model, unlike in standard FD method, RBF-FD need not to
  use the maximum velocity and consequently over-sampled nodes in some parts of the domain. This
  lowers the total number of required nodes for highly-complicated velocity models.
  \item Although RBF-FD can theoretically deal with highly-non uniform
  node-distributions, the non-uniformity introduces numerical dispersion. However,
  since this error is mostly near the source, its contribution to the final observation is not as
  noticeable as the corresponding undersampled FD method. 
  \item This manuscript provides the first high-performance (C++) open-source repository for meshfree seismic modeling. We believe the source-codes of the present paper will help readers to have a better understanding of state-of-the art implementation of RBF-FD and promote further improvement in the field with minimal development efforts.
\end{enumerate}

% The paper presents a meshfree (RBF-FD) solution of time-domain acoustic wave equation for seismic forward modeling where the meshfree nodes are adaptively refined with regards to the velocity model. 
% We show that the RBF-FD can be used with Absorbing Boundary Conditions (ABC), without any special treatment of nodes near the boundaries. 
% In the revised version of the manuscript, we have shown a modeling example for Canadian Foothill Model, which has a non-smooth irregular topography.  
% The state-of-the-art RBF-FD method is getting popular and being explored more and more in different fields. The numerical analysis presented in this paper adds towards the robustness of the RBF-FD method. 
% We also discuss a comparison of RBF-FD (a meshfree finite-difference method) to standard mesh-based finite difference method. 
% This manuscript provides the first high-performance (C++) open-source repository for meshfree seismic modeling 
\section*{Author's Contribution}
Gregor Kosec and Jure Slak prepared the approximation engines used for numerical solution of the considered problem
Jure M. Berljavac, and Pankaj K Mishra prepared numerical solution procedure and analysed results. All authors equally contributed in preparation of the manuscript. 

\section*{Computer Code Availability}
The computer codes and instruction to reproduce the results in this paper are freely available at \url{https://gitlab.com/e62Lab/2019_p_wavepropagation_code} .
\bibliographystyle{unsrtnat}
\bibliography{references}
\end{document}